\RequirePackage{fix-cm}
 \documentclass[twocolumn]{svjour3}          % twocolumn
\smartqed  % flush right qed marks, e.g. at end of proof
\usepackage{multirow}
\usepackage{graphicx}
\usepackage{subcaption}
\usepackage{amsmath} 
\usepackage{tikz}
\def\checkmark{\tikz\fill[scale=0.4](0,.35) -- (.25,0) -- (1,.7) -- (.25,.15) -- cycle;} 

\usepackage{bbm} 

\usepackage{xcolor}

\newcommand{\squishlist}{
   \begin{list}{$\bullet$}
    { \setlength{\itemsep}{0pt}      \setlength{\parsep}{3pt}
      \setlength{\topsep}{3pt}       \setlength{\partopsep}{0pt}
      \setlength{\leftmargin}{1.5em} \setlength{\labelwidth}{1em}
      \setlength{\labelsep}{0.5em} } }

\newcommand{\squishend}{
    \end{list}  }

\begin{document}

\title{Fairness in Rankings and Recommendations: An Overview}

\author{Evaggelia Pitoura  \and Kostas Stefanidis \and Georgia Koutrika}


\institute{Evaggelia Pitoura \at
              University of Ioannina, Greece \\
              Tel.: +30-26510-08811\\
              \email{pitoura@cs.uoi.gr}       
           \and
          Kostas Stefanidis * \at 
                Tampere University, Finland \\
              Tel.: +358-50-4174121\\
              \email{konstantinos.stefanidis@tuni.fi}         
    \and
        Georgia Koutrika \at
        Athena Research Center, Greece \\
              Tel.: +30-210-6875351\\
              \email{georgia@athenarc.gr}    
}

\date{Received: date / Accepted: date}

\maketitle

\begin{abstract}
We increasingly depend on a variety of data-driven algorithmic systems to assist us in many aspects of life. Search engines and recommender systems amongst others are used as sources of information and to help us in making all sort of decisions from selecting restaurants and books, to choosing friends and careers. This has given rise to important concerns regarding the fairness of such systems. In this work, we aim at presenting a toolkit of definitions, models and methods used for ensuring fairness in rankings and recommendations. Our objectives are three-fold: (\emph{a}) to provide a solid framework on a novel, quickly evolving, and impactful domain, (\emph{b}) to present related methods and put them into perspective, and (\emph{c}) to highlight open challenges and research paths for future work.

\end{abstract}

\section{Introduction}
\label{intro} 
Algorithmic systems, driven by large amounts of data, are increasingly being used in all aspects of society to assist people in forming opinions and taking decisions. Such algorithmic systems offer enormous opportunities, since they accelerate scientific discovery in various domains, including personalized medicine, smart weather forecasting and many other fields. They can also automate tasks regarding simple personal decisions, and help in improving our daily life through personal assistants and recommendations, like where to eat and what are the news. Moving forward, they have the potential of transforming society through open government and many more benefits. 

Often, such systems are used to assist, or, even replace human decision making in diverse domains. Examples include software systems used in school admissions, housing, pricing of goods and services, credit score estimation, job applicant selection, and sentencing decisions in courts and surveillance. This automation raises concerns about how much we can trust such systems. 

%what images do people choose to represent careers, like for instance, 

A steady stream of studies  has shown that decision support systems can unintentionally both encode existing human biases and introduce new ones \cite{DBLP:journals/cacm/ChouldechovaR20}.
For example, in image search, when the query is  about doctors or nurses, what is the percentage of images portraying women that we get in the result? Evidence shows stereotype exaggeration and  systematic underrepresentation of women when compared with the actual percentage, as estimated by the US Bureau of labor and statistics \cite{DBLP:conf/chi/KayMM15}.
Two interesting conclusions from the study were that people prefer and rate search results higher when these results are consistent with stereotypes. Another interesting result is that if you shift the representation of gender in image search results then the people's perception about real world distribution tends to shift too.

Another well-known example is the COMPAS system, which is a commercial tool that uses a risk assessment algorithm to predict some categories of future crime. Specifically, this tool is used in courts in the US to assist bail and sentencing decisions, and it was found that the false positive rate, that is the people who were labeled by the tool as high risk but did not re-offend, was nearly twice as high for African-American as for white defendants \cite{Angwin16}. This means that many times the ubiquitous use of decision support systems may create possible threats of economic loss, social stigmatization, or even loss of liberty. 
There are many more case studies, like the above ones. For example, names that are used by men and women of color are much more likely to generate ads related to arrest records \cite{DBLP:journals/cacm/Sweeney13}. Also using a tool called Adfisher, it was found that if you set the gender to female, this will result in getting ads for less high paid jobs\footnote{https://fairlyaccountable.org/adfisher/}. Or, in the case of word embeddings the vector that represents computer programming is closer to men than to women.

Data-driven systems are also being employed by search and recommendation engines in movie and music platforms, advertisements, social media, and news outlets, among others. Recent studies report that social media has become the main source of  online news with more than 2.4 billion internet users, of which nearly 64.5\% receive breaking news from social media instead of traditional sources \cite{martin18}. Thus, to a great extent, search and recommendation engines in such systems play a central role in shaping our experiences and influencing our perception of the world.

For example, people come to their musical tastes in all kinds of ways, but how most of us listen to music now offers specific problems of embedded bias. When a streaming service offers music recommendations, it does so by studying what music has been listened to before.  That creates a suggestions loop, amplifying existing bias and reducing diversity.  A recent study 
analysed the publicly available listening records of 330,000 users of one service and showed that female artists only represented 25 per cent of the music listened to by users. The study identified that gender fairness is one of the artists' main concerns as female artists are not given equal exposure in music recommendations \cite{DBLP:conf/chiir/FerraroSB21}.

%\textcolor{blue}{ Besides serving diverse groups of users, they also need to represent and serve item providers fairly as well. In interviews with music artists, .}

Another study led by University of Southern California on Facebook ad recommendations revealed that the recommendation system disproportionately showed certain types of job ads to men and women \cite{Imana21a}.
The system was more likely to present job ads to users if their gender identity reflected the concentration of that gender in a particular position or industry. Hence, recommendations amplified existing bias and created fewer opportunities for people based on their gender. However, ads may be targeted based
on qualifications, but not on protected categories, based on US law.

%There are many reports questioning the output of such systems. For instance, a known study on search results showed evidence for stereotype exaggeration in images returned when people search for professional careers \cite{DBLP:conf/chi/KayMM15}.

In this article, we pay special attention to the concept of fairness in rankings and recommendations. By fairness, we typically mean lack of discrimination (bias). Bias may come from the algorithm, reflecting, for example, commercial or other preferences of its designers, or even from the actual data, for example, if a survey contains biased questions, or, if some specific population is misrepresented in the input data.

As fairness is an elusive concept, an abundance of definitions and models of fairness have been proposed as well as several algorithmic approaches for fair rankings and recommendations making the landscape very convoluted. In order to make real progress in building fair-aware systems, we need to de-mystify what has been done, understand how and when each model and approach can be used, and, finally, distinguish the research challenges ahead of us. 

Therefore, we follow a systematic and structured approach to explain the various sides of and approaches to fairness.
In this survey, we present fairness models for rankings and recommendations  separately from the  computational methods used to enforce them, 
since  many of the computational methods originally introduced for a specific model  are applicable to other models as well.  By providing an overview of the spectrum of different models and computational methods, new ways to combine them may evolve. 

We start by presenting fairness models.
First, we provide a birds' eye view of how notions of fairness in rankings and recommendations have been formalized.
 We also present a taxonomy. Specifically, we distinguish between \textit{individual} and \textit{group} fairness, \textit{consumer} and \textit{producer} fairness, and fairness for \textit{single} and \textit{multiple} outputs. 
Then, we present concrete models and definitions for rankings and recommendations. We highlight their differences and commonalities, and present how these models fit into our taxonomy.
%Then, we present concrete models and definitions for rankings, recommendations and rank aggregation

We describe solutions for fair rankings and recommendations. We organize them into \emph{pre-processing approaches} that aim at transforming the data to remove any underlying bias or discrimination, \emph{in-processing approaches} that aim at modifying existing or introducing new algorithms that result in fair rankings and recommendations, and \emph{post-processing approaches} that modify the output of the algorithm. Within each category, we further classify approaches along several dimensions. 
We discuss other cases where a system needs to make decisions and where fairness is also important, and present open research challenges pertaining to fairness in the broader context of data management.

To the best of our knowledge, this is the first survey that provides a toolkit of definitions, models and methods used for ensuring fairness in rankings and recommendations.
A recent survey focuses on fairness in ranking \cite{DBLP:journals/corr/abs-2103-14000}.
The two surveys are complementary to each other both in terms of perspective and in terms of coverage.
Fairness is an evasive concept and integrating it in algorithms and systems is an emerging fast-changing field. We provide a more technical classification of recent work, whereas the view in \cite{DBLP:journals/corr/abs-2103-14000} is a socio-technical one that aims at placing the various approaches to fairness within a value framework. Content is different as well, since we  also cover  recommendations and rank aggregation.
%
%{\color{blue} At the same time, \cite{DBLP:journals/corr/abs-2103-14000} presents a selection of approaches for fairness only in rankings. In contrast, our survey covers as well fairness in recommender systems and rank aggregation, offering a broader perspective, connecting works 
%across different fields. The overlap of the two surveys is kept to the minimum. }
Recent tutorials, with a stricter focus than ours, focusing on concepts and metrics of fairness and the challenges in applying these to recommendations and information retrieval, as well as to scoring methods, are presented, respectively, in \cite{DBLP:conf/sigir/EkstrandBD19} and \cite{DBLP:journals/pvldb/AsudehJ20,DBLP:conf/www/OosterhuisJR20}. 
On the other hand, this article has a much wider coverage and depth, presenting a structured survey and comparison of methods and models for ensuring fairness in rankings and recommendations. 

The remaining of this survey is organized as follows. Section 2 presents the core definitions of fairness, and Section 3 reviews definitions of fairness that are applicable specifically to rankings, recommenders and rank aggregation. Section 4 discusses a distinction of the methods for achieving fairness, while Sections 5, 6 and 7 organize and present in detail the pre-, in- and post-processing methods. Section 8 offers a comparison between the in- and post-processing methods. 
Section 9 studies how we can verify whether a program is fair. Finally, Section \ref{sec:challenges} elaborates on critical open issues and challenges for future work, and Section \ref{sec:conclusions} summarizes the status of the current research on fairness in ranking and recommender systems.

\section{The Fairness Problem}
\label{sec:fair-problem}
In this section, we  start with an overview of approaches to modeling fairness and  then provide a taxonomy of the different types of fairness models in ranking and recommendations.

\subsection{A General View on Fairness}
Most approaches to algorithmic fairness interpret fairness as   \textit{lack of discrimination} \cite{DBLP:journals/corr/FriedlerSV16,DBLP:journals/cacm/FriedlerSV21}, asking that  an algorithm should not discriminate against its input entities based on attributes that are not relevant to the task at hand.
Such attributes are called \textit{protected}, or \textit{sensitive}, and often include among others gender, religion, age, sexual orientation, and race.

So far, most work on defining, detecting and removing unfairness has focused on classification algorithms used in decision making.
In \textit{classification algorithms}, each input entity is assigned to one from a set of predefined classes. In this case, characteristics of the input entities that are not relevant to the task at hand
should not influence the output of the classifier.
For example, the values of protected attributes should not hinder the assignment of an entity to the positive class, where the positive class may 
for example, correspond to getting a job, or, being admitted at a school.

In this paper, we focus on ranking and recommendation algorithms.
Given a set of entities, $\{i_1, i_2, \dots i_N\}$, a \textit{ranking algorithm} produces a ranking $r$ of the entities, where $r$ is an assignment (mapping) of entities to ranking positions. Ranking is based on some measure of the relative quality of the entities for the task at hand.
For example, the entities in the output of a search query are ranked mainly based on their relevance to the query.
In the following, we will  also refer to the measure of quality, as  \textit{utility}.
Abstractly, a fair ranking is one where the assignment of entities to positions is
not unjustifiably influenced by the values of their protected attributes.

\textit{Recommendation systems} retrieve interesting items for users based on their profiles and their  history. Depending on the application and the recommendation system, history may include explicit user ratings of items, or, selection of items (e.g., views, clicks). 
In general, recommenders estimate a  score, or, rating,
$\hat{s} (u, i)$ for  a user  $u$ and an item $i$  that reflects the preference of user $u$ for item $i$, or, in other words, the relevance of item $i$ for user $u$. Then, a recommendation list $I$ is formed for user $u$ that includes the  items having the highest estimated score for  $u$. These scores can be seen as the utility scores in the case of recommenders.
In abstract terms, a recommendation is fair, if the values of the protected attributes of the users, or, the items, do not affect the 
outcome of the recommendation.

On a high level, we can distinguish between two approaches to formalizing fairness \cite{DHP+12}:
\begin{itemize}
	\item 
	\textit{Individual fairness} definitions are based on the premise that similar entities should be treated similarly.
	\item 
	\textit{Group fairness} definitions group entities based on the value of one or more protected attributes and ask that 
	all groups are treated similarly.
\end{itemize}

To operationalize both approaches to fairness, we  need to define similarity for the input and the output of an algorithm. 
For \emph{input similarity}, we need a means of quantifying  similarity of entities in the case of individual fairness, and, a way of partitioning entities into groups, in the case of group fairness. For \emph{output similarity}, for both individual and group fairness, we need  a formal definition of what similar treatment means.

\vspace*{0.10in} 
\noindent \textbf{Input similarity.} 
For individual fairness, a common approach to defining input similarity is a distance-based one \cite{DHP+12}.
Let  $V$ be the set of entities, we assume there is a distance metric $d: V \times V \rightarrow R$ between each pair of entities, such that, the more dissimilar the entities, the larger their distance.
This metric should be task-specific, that is, two entities may be similar for one task and dissimilar for another. For example,
two individuals may be consider similar to each other (e.g., have similar qualifications) when it comes to being
admitted to college but dissimilar
when it comes to receiving a loan.
The metric may be externally imposed, e.g., by a regulatory body, or externally proposed, e.g., by a civil rights organization.
Ideally, the metric should express the ground truth, or, the best available approximation of it. Finally,  this metric should be
made public, and open to discussion and refinement.
For group fairness, the challenge lies in 
determining how  to partition entities  into  
groups. 
%Often looking only at the protected attributes does not suffice, since there may be other 
%\textit{proxy} attributes correlated with the protected ones, a case also known as 
%%\textit{redundant encoding}.
%%++ Add example with zip codes here ++

\vspace*{0.1in} \noindent
\textbf{Output similarity.}
Specifying what it means for entities, or groups of entities to be treated similarly is an intricate problem, from both a social and a technical perspective.
From a social perspective, a fundamental distinction is made
between equity and equality.
Simply put, \textit{equality} refers to treating entities equally, while \textit{equity}
refers to treating entities according to their needs, so that they all finally receive the same output, even when some individuals are disadvantaged.  

Note that, often \textit{blindness}, i.e., hiding the values of the protected attributes, does not suffice to produce fair outputs, since there may be other 
\textit{proxy} attributes correlated with the protected ones, a case also known as 
\textit{redundant encoding}.
Take for example a  zip code attribute. Zip codes may reveal sensitive information when the majority of the residents of a neighborhood belong to a specific ethnic group \cite{redlining}. 
In fact, such considerations have led to making \textit{redlining}, i.e., the practice of arbitrary denying or limiting financial services to specific neighborhoods, illegal in the US.

Another social-based differentiation can be made between 
disparate treatment and disparate impact.
\textit{Disparate treatment} is the often illegal practice of treating an entity differently based on its protected attributes.
\textit{Disparate impact} refers to cases where the output depends on 
the protected attributes, even if all entities are treated the same way.
The disparate impact doctrine  was solidified in the US after [Griggs v. Duke Power Co. 1971] where a high school diploma was required for unskilled work, excluding  applicants of color.

From a technical perspective, how output similarity is translated into
quantifiable measures depends clearly on the specific type of algorithm. 
In this paper, we focus on ranking and recommendation algorithms.

Overall, in the case of ranking, a central issue is the manifestation of
\textit{position bias}, i.e., the fact that people tend to consider only the items that appear in the top few positions of a ranking.
Even more the attention that items  receive, that is the visibility of the items, is highly skewed with regards to their position in the list.
At a high level, output similarity in the case of ranking  refers to offering similar visibility to similar items or group of items, that is, placing them at similar positions in the ranking, especially when it comes to top positions.

For recommendations, one approach to defining output fairness  is to consider the recommendation problem as a classification problem where the positive class is the recommendation list.
Another approach is to consider the recommendation list $I$ as a ranked list $r$, in which case, the position of each recommended item in the list should also be taken into account.

We  refine individual and group fairness based on the type of output similarity in the next section.

\subsection{A Taxonomy of Fairness Definitions}
\begin{figure*}[t]
\begin{center}
\includegraphics[width=1.7\columnwidth]{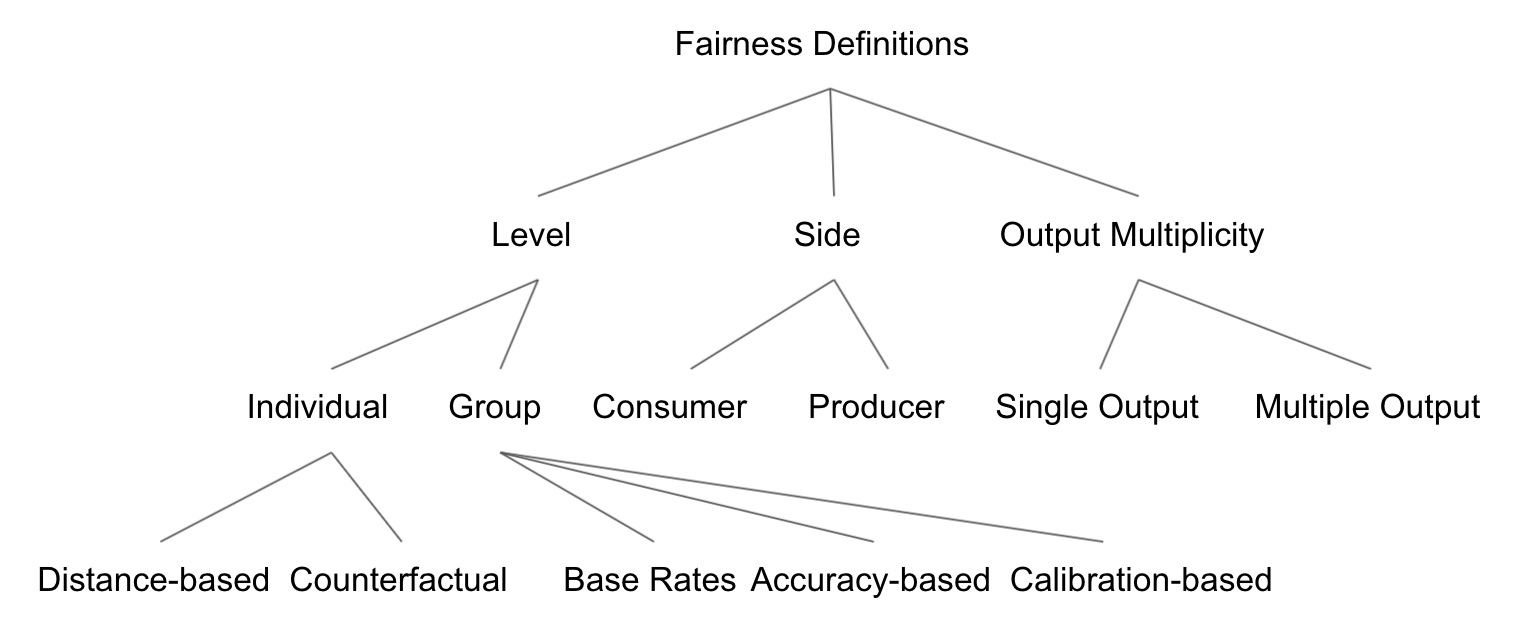}
\caption{Fairness Definitions Taxonomy.} \label{fig:multisidedtaxonomy}       
\end{center}
\end{figure*}

In the previous section, we distinguished between  individual and group fairness formulations. We will refer to this distinction of fairness models as  the \textit{level} of fairness. When it comes to ranking and recommendation systems, besides different levels, we also have more than one \textit{side} at which fairness criteria are applicable. Finally, we differentiate fairness models based on whether the fairness requirements are applied at a single or at \textit {multiple outputs} of an algorithm.
%+++ may also refer to dynamic: not knowing the whole input? +++
The different  dimensions of fairness are summarized in Figure \ref{fig:multisidedtaxonomy}.

Note, that the various fairness definitions in this and the following sections can be used both: (a) as conditions that a system must satisfy for being fair, and (b) as measures of fairness.
For instance, we can measure how much a fairness condition is violated, or, define a condition by setting a threshold on a fairness measure. 

\subsubsection{Levels of fairness} \label{sec:levels}

We now refine individual and group fairness based on how output similarity is specified.
Seminal research in formalizing algorithmic fairness has focused on
classification algorithms used in decision making. Such research has been influential in 
the study of fairness in other types of algorithms. So, we refer to such research briefly here and relate it to ranking and recommendations. 

\vspace*{0.15in}
\noindent \textbf{Types of Individual Fairness.} 
One way of formulating individual fairness is a distance-based one. Intuitively, given a distance measure $d$ between two entities, and a distance measure $D$ between the outputs of an algorithm, we would like the distance between the output of the algorithm for two entities to be small, when the entities are similar. Let us see a concrete example from the area of probabilistic classifiers \cite{DHP+12}.

Let $M$ be a classifier that maps entities $V$ to outcomes $A$. In the case of probabilistic classifiers, 
these are randomized mappings from entities to probability distributions over outcomes. 
Specifically, to classify an entity $v \in V$, we choose an outcome $a \in A$ according to the distribution $M(v)$. 
We say that a classifier is individually fair if the mapping $M:$ $V \rightarrow \Delta(A)$ 
satisfies the $(D, d)$-Lipschitz property,  that is, $\forall \, v, u \in V$,
$D(M(v), M(u)) \leq d(v, u)$,
where $D$ is a distance measure between probability distributions, and $d$ a distance metric between entities. In words, the distance $D$  between probability distributions assigned by the classifier $M$ should be no greater than the actual distance $d$ between the entities.

Another form of individual fairness is \textit{counterfactual fairness} \cite{KLR+17}.
The intuition in this case, is that an output is fair towards an entity if it is the same in both the actual world and a counterfactual world where the entity belonged to a different group. Causal inference is used to formalize this notion of fairness.

\vspace*{0.15in}
\noindent{\bf Types of Group Fairness.} 
For simplicity, let us assume two groups, namely,  the protected group $G^+$ and the non-protected (or, privileged) group $G^-$.
We will start by presenting statistical approaches commonly used in classification.
Assume that $Y$ is the actual and $\hat{Y}$ the predicted output of a binary classifier, that is, $Y$ is the ground truth, and  $\hat{Y}$ the output of the algorithm.
Let $1$ be the positive class that leads to a favorable decision, e.g., someone getting a loan, or being admitted at a competitive school, and $S$ be the predicted probability for a certain classification.

Statistical approaches to group fairness can be distinguished as \cite{FSV+19,VR18}:
\begin{itemize}
\item \textit{base rates approaches}: that use only the output $\hat{Y}$ of the algorithm, 
 \item \textit{accuracy approaches}: that use both the output $\hat{Y}$ of the algorithm and the ground truth $Y$, and 
 \item \textit{calibration approaches}: that use the predicted probability $S$ and the ground truth $Y$.
 \end{itemize}

In classification, \textit{base rate fairness} compares the probability $P(\hat{Y}=1|v \in G^+)$ that an entity $v$  receives the favorable outcome when $v$ belongs to the protected group $G^+$ with the corresponding probability $P(\hat{Y}=1|v \in G^-)$ that $v$ receives the favorable outcome when $v$ belongs to the non-protected group $G^-$. 
To compare the two, we may take their ratio \cite{ZWS+13}, \cite{FFM+15}: 
%\begin{equation}
$\frac{P(\hat{Y}=1|v \in G^+)}{P(\hat{Y}=1|v \in G^-)}$
%\end{equation} 
or, their difference \cite{CV10}:
%\begin{equation}
$1 - (P(\hat{Y}=1|v \in G^+) -P(\hat{Y}=1|v \in G^-))$.
%\end{equation} 

Now, in abstract terms, a base rate fairness definition for ranking may compare the probabilities of items
from each group to appear in similarly good ranking
positions, while for recommendations, the probabilities
of them being recommended.

When the probabilities of a favorable outcome are equal for the two groups, we have a special type of fairness termed {\textit demographic}, or  \textit{statistical parity}. 
Statistical parity preserves the input ratio, that is, the demographics of the individuals receiving a favorable outcome are the same as the demographics of the underlying population.
Statistical parity is a natural way to model equity: members of each group have the
same chance of receiving the favorable output.

Base rate fairness ignores the actual output, the output may
be fair, but it may not reflect the ground truth. For example, assume that the classification task is getting or not a job and the protected attribute is 
gender. Statistical parity asks for a specific ratio of women in the positive class, even when
there are not that many women in the input who are well-qualified for the job. 
Accuracy and calibration look at  traditional evaluation measures and require that the algorithm works equally well in terms of prediction errors for both groups.

%Based on a generalization of the 80 percent rule advocated by the US Equal Employment Opportunity Commission, the case of the ration  being smaller than a threshold $\tau = 0.8$, is considered a case of disparate impact (unintended discrimination) \cite{FFM+15}. 

In classification, \textit{accuracy-based fairness} warrants that various types of  classification errors (e.g., true positives, false positives) are equal across groups.
Depending on the type of classification errors considered, the achieved type of fairness takes
different names \cite{HPS16}.
For example, the case in which, we ask that $P(\hat{Y}=1| Y = 1, v \in G^+)$ = $P(\hat{Y}=1|Y =1, v \in G^-)$ (i.e., the case of equal true positive rate for the two groups) is called \textit{equal opportunity}.

Comparing equal opportunity with statistical parity, 
again the members of the two groups have the same chance of getting the favorable outcome, but  only when these members  qualify.
Thus, equal opportunity is more close to an equality interpretation of fairness.

In analogy to accuracy,  in the case of ranking, the ground truth is reflected in the utility of the items. Thus, approaches that take into account utility when defining fairness in ranking can be seen as accuracy-based ones. In recommendations, accuracy-based definitions look at the differences between the actual and the predicted ratings of the items for the two groups. 

\textit{Calibration-based} fairness considers probabilistic classifiers
that predict a probability for each class \cite{Ch17,KMR17}.
In general, a classification algorithm is considered to be well-calibrated if: when  the algorithm predicts a set of individuals as having probability $p$ of  belonging to the positive class,
 then approximately a $p$ fraction of this set are actual members  of the positive class.
%the predicted proportions of the various classes agree with the actual proportions of the data items in the input data.
In terms of fairness, intuitively, we would like the classifier to be equally well-calibrated for both groups.
An example calibration-based fairness is asking that for any predicted probability score $p$ in $[0, 1]$, the probability of actually getting a favorable outcome
is equal for both groups, i.e.,
$P(Y=1| S = p, v \in G^+)$ = $P(Y=1|S = p, v \in G^-)$.

Group-based measures in general tend to ignore the merits of each individual in the group. Some individuals in a group may be better for a given task than other individuals in the group, which is not
captured by some group-based fairness definitions. This issue may lead to two problematic behaviors, namely, (a) the
\textit{self-fulfilling prophecy} where by
deliberately choosing the less qualified members of the protected group we aim at building a bad track record for the group
and (b) \textit{reverse tokenism} where by
not choosing a well qualified member of the non-protected group
we aim at creating convincing refutations for the members of the protected group that are also not selected.

\subsubsection{Multi-sided fairness} 
In ranking and recommendations, there are at least two sides involved: the items that are being ranked, or recommended, and the users that receive the rankings or the recommendations. We distinguish between \textit{producer} or \textit{item-side} fairness and \textit{consumer} or \textit{user-side} fairness. 
Note that the items that are being ranked or recommended may be also people, for example, in case of ranking job applicants, but we call them items for simplicity. 

\textit{Producer} or \textit{item-side} fairness focuses on the items that are being ranked, or recommended.
In this case, we would like similar items or groups of items to be ranked, or, be recommended in a similar way, e.g., to appear in similar positions in a ranking. 
This is the main type of fairness, we have discussed so far.
For instance, if we consider political orientation as the protected attribute of an article, we may ask that the value of this attribute does not affect the ranking of articles in a search result, or a news feed.

\textit{Consumer} or \textit{user-side} fairness focuses on the users who receive, or consume the data items in a ranking, e.g., a search result, or a recommendation.
In abstract terms, we would like similar users, or 
groups of users, to receive similar rankings or recommendations. For instance, if gender is the protected attribute of a user receiving job recommendations, we may ask that the gender of the user does not influence the job recommendations that the user receives. 

%When comparing consumer and producer fairness, we note that the latter is passive, since typically producers do not seek suggestion opportunities that cover their information needs, but instead just wait for the users to ask for suggestions. 
There are cases in which a system may require fairness for both consumers and providers, when for instance both the users and the items belong to protected groups. 
For example, assume a rental property business that 
 wishes to treat minority applicants as a protected class and ensure that they have access to properties similar to other renters, 
while at the same time, wishes to treat minority landlords as a protected class and ensure that highly qualified tenants are referred to them at the same rate as to other landlords.

Different types of recommendation systems may call for specializations of consumer and producer fairness.
Such a case is group recommendation systems. Group recommendation systems  recommend items to groups of users as opposed to a single user, 
for example a movie to a group of friends, an event to an online community, or a excursion to a group of tourists \cite{grouprec1,grouprec2}.
In this case, we  have  different types of  consumer fairness, since now the consumer is not just a single user.
Another case are bundle and package recommendation systems that recommend complex items, or sets of items, instead of just a single item, for example a set of places to visit, or courses to attend \cite{packagerec1}.
In this case, we may have different types of producer fairness, since now the recommended items are composite. 
We discuss these special types of fairness in Section \ref{sec:group-rec-fairness-defs}.

We can expand the sides of fairness  further by  considering the additional stakeholders that may be involved in a recommendation system besides the consumers and the producers. For example, in a recommendation system, the items being recommended may belong to different providers. For instance, in the case of movie recommendations, the movies may be produced by different
studios. In this case, we may ask for producer fairness with respect to the \textit{providers} of the items, instead of the single items. For example, in an online craft marketplace, we may want to ensure market diversity and avoid monopoly domination,  where the system wishes to ensure that new entrants to the market get a reasonable share of recommendations even though they have fewer shoppers than established vendors. 
Note that one way to model provider fairness is by treating the provider as a protected attribute of the items. 

%This may also be interpreted as group fairness with the protected attribute being the provider.

Other sides of fairness include: (a) fairness for the owners of the recommendation system, especially when the owners are different than the producers,
and (b) fairness for system regulators and auditors,  for example, data scientists, machine learning researchers, policymakers and governmental auditors that are using the system for decision making. 

\subsubsection{Output Multiplicity} 
There are may be cases in which it is not feasible to achieve fairness by considering just a single ranking or a single recommendation output.
Thus, recently, there exist  approaches that propose achieving fairness via a series of outputs.
Consider, for example, the case where the same items appear in the results of multiple search queries, or the same users receive more than one recommendation. In such cases, we may ask that an item is not necessarily being treated fairly in each and every search result but the item is treated fairly overall in a set of a search results. Similarly, a  user may be treated 
unfairly in a single recommendation but fairly in a sequence of recommendations.

Concretely, we distinguish between \textit{single output} and
\textit{multiple output} fairness. 
In multiple output fairness, we ask for \textit{eventual}, or \textit{amortized} consumer, or producer fairness, i.e., we ask that the consumers or producers are treated fairly in a series of rankings or recommendations as a whole, although they may be treated unfairly in one or more single ranking or recommendation in the series.  
%In the case of a \emph{single output} of the system, i.e., a ranking or recommendation,  we seek to define fairness in this context. 
%In the case of \emph{multiple output} fairness, we consider a sequence of rankings, or recommendations. For example, in the case where the same items appear in results of multiple search queries, or the same users receive more than one recommendation. In this case, we may ask for \textit{eventual}, or \textit{amortized} consumer, or producer fairness, i.e., we ask that the consumers or producers are treated fairly in the sequence of rankings of recommendations as a whole, although they may be treated unfairly in a single ranking or recommendation.

Another case is \textit{sequential recommenders} that suggest items of interest by modeling the sequential dependencies over the user-item interactions in a sequence. This means that the recommender treats the user-item interactions as a dynamic sequence and takes the sequential dependencies into account to capture the current and previous user preferences for increasing the quality of recommendations \cite{DBLP:conf/sac/StratigiNPS20}. The system recommends different items at each interaction, while retaining knowledge from past interactions.
Interestingly, due to the multiple user-item interactions in sequential recommender systems,  fairness correction can be performed, while moving from one interaction to the next. 

\section{Models of Fairness}
\label{sec:fair-rank-recs}
In the previous section, we presented an overview of fairness and its various dimensions in ranking and recommendations. In this section, we  present a number of concrete models and definitions of fairness proposed for ranking, recommendations and rank aggregation.

\subsection{Fairness in Rankings}  \label{sec:fair-rank}
Most approaches to fairness in ranking handle producer fairness, that is, their goal is to ensure that the items being ranked are treated fairly.
In general, ranking fairness asks that similar items or group of items receive similar visibility, that is, they appear at similar positions in the ranking.
A main issue  is accounting for position bias.
Since in western cultures, we read from top to bottom and from left to right, the visibility of lower-ranked items drops
rapidly when compared to the visibility of  higher-ranked ones \cite{ricardo}.

We will use the example rankings in Figure \ref{fig:ranking}. The protected attribute is color, red is the protected group and score is the utility of the item. The ranking on the left ($r_l$) corresponds to a ranking based solely on utility, the ranking in the middle ($r_m$) achieves the highest possible representation of the protected group in the top positions, while the ranking in the right ($r_r$) is an intermediate one.

\begin{figure*}
	\centering
	\begin{subfigure}[t]{0.3\textwidth} \centering
		\includegraphics[scale=0.5]{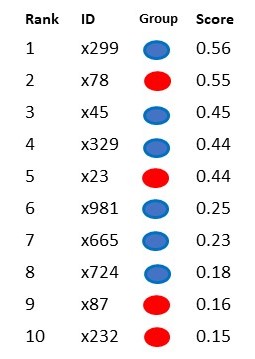}
		\caption{$r_l$}
	\end{subfigure}%\hspace{0.1\textwidth}
	\begin{subfigure}[t]{0.3\textwidth} \centering
		\includegraphics[scale=0.5]{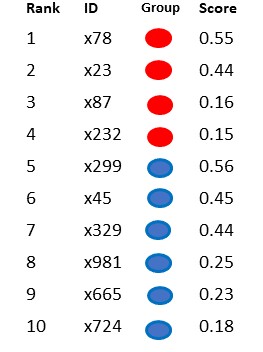}
		\caption{$r_m$}
	\end{subfigure}
	\begin{subfigure}[t]{0.3\textwidth} \centering
		\includegraphics[scale=0.5]{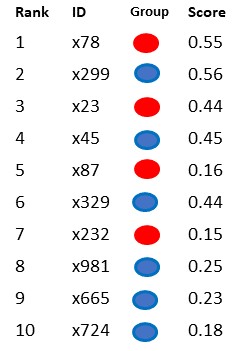}
		\caption{$r_r$}
	\end{subfigure}
\label{fig:ranking} 
	\caption{Example rankings: (a) $r_l$ is based solely on utility, (b) $r_m$ is the optimal ranking in terms of the representation of the protected group,  (c) $r_r$ is an intermediate ranking between the two. Red is the protected value.}\label{fig:ranking} 
	%\vspace{-0.25in}
\end{figure*}

\textbf{Fairness constraints}.
A number of group-based fairness models for ranking focus on the representation  (i.e., number of items) of the 
protected group in the top-$k$ position in the ranking.
One such type of group fairness is achieved by
constraining the number of items from the different groups that can  appear in the top-$k$ positions. Specifically, in the
\textit{fairness constraints} approach \cite{CSV18},  given a number of protected attributes, or, properties, fairness requirements are expressed by specifying
an upper bound $U_{l,k}$ and  a lower bound $L_{l,k}$ on the number of items with property $l$ that are allowed to appear in the top-$k$ positions of the ranking. 
For example, in Figure \ref{fig:ranking}, the fairness constraint $L_{red, 4} = 2$, that requires that there are at least 2 items with property red in the top-4 positions, is satisfied by rankings $r_m$ and $r_r$, but not by ranking $r_l$.

\textbf{Discounted cumulative fairness}. 
Another approach looks at the proportional representation of the items of the protected group at top-$p$ prefixes of the ranking for various values of $p$ \cite{YS17}.
The proposed model builds on  the Discounted Cumulative Gain (DCG) measure.
DCG is a standard way of  measuring the ranking quality of the top-$k$ items. $DCG@k$  accumulates the utility, $util(j)$, of each item at position $j$ from the top position up to position $k$ logarithmically discounted by the position $j$ of the item, thus favoring higher utility scores at first positions: 
\begin{equation}
DCG@k(r) = \sum_{j=1}^k \frac{util(j)}{log_2 (j+1)}, 
\end{equation} 
The DCG value is then normalized by the DCG of the perfect ranking for obtaining NDCG. 
For example, the $DCG$ of the three ranking in Figure \ref{fig:ranking} are $DCG@10(r_l)$ $ = 1.81$,  $DCG@10(r_m)$ $= 1.7$ and $DCG@10(r_r)$ $ = 1.77$. Clearly, $r_l$ that ranks items solely by utility has the largest $DCG$ value. 

\textit{Discounted cumulative fairness} accumulates the number of items belonging to the protected group $G^+$ at discrete positions in the ranking (e.g., at positions $p$ = 5, 10, …) and  discounts these numbers accordingly, so as to favor the representation of the protected group at prefixes at higher positions. 
Three different definitions based on this general idea have been provided.

The first one, the \textit{normalized discounted difference} ($rND$) of a ranking $r$, measures the difference in the proportion of the items of the protected group in the top-$p$ prefixes, for various values of $p$, and in the overall population:
\begin{equation}
\centering
rND(r) = \frac{1}{opt\_rND} \sum_{p = 5, 10, \dots}^N \frac{1}{log_2(p)}|\frac{|G_{1 \dots p}^+|}{p} - \frac{|G^+|}{N}|
\label{eq:rndr}
\end{equation}
where $N$ is the total number of items, $|G_{1 \dots p}^+|$ is the number of items of the protected group in the top-$p$ positions  and  $opt\_rND$ the optimal value.

For example, the optimal value in Figure \ref{fig:ranking} is the one of ranking $r_m$, that is  of the ranking with the maximum possible representation for the protected group and it is equal to
$opt\_rND(r)$ $= \frac{1}{log_2(5)} |\frac{4}{5} - \frac{4}{10}|$ = 0.93.
The $r_m$ ranking which is based on utility has the smallesr $rND$:
$ rND(r_m)$ $ = \frac{1}{0.93}
(  (\frac{1}{log_2(5)}|\frac{2}{5} - 
\frac{4}{10}| + 
\frac{1}{log_2(10)}|\frac{4}{10} - 
\frac{4}{10}|) = 0
$,
while for $r_r$  we have:
$ rND(r_r)$ $ = \frac{1}{0.93}
(  (\frac{1}{log_2(5)}|\frac{3}{5} - 
\frac{4}{10}| + 
\frac{1}{log_2(10)}|\frac{4}{10} - 
\frac{4}{10}|) = 0.5.
$

A variation, termed \textit{normalized discounted ratio} measures the difference between the proportion of the items of the protected group in the top-$p$ positions and the items of the non-protected group in the top-$p$ positions, for various values of $p$.
This is achieved by modifying
 Equation \ref{eq:rndr} so that instead of dividing  with $p$, i.e., the total number of items up to position $p$, we divide with
$ |G_{1 \dots p}^-|$ (i.e., the number of items of the protected group in the top-$p$ positions)
and instead of dividing with $N$, we divide by
$|G^-|$. 

Finally, the \textit{normalized KL divergence} ($rKL$) definition of fairness
uses KL-divergence to compute the expectation of the difference between the membership probability distribution of  the protected group at the top-$p$ positions (for $p$ = 5, 10, $\dots$) and in the overall population.

\textbf{Fairness of exposure}. A problem with the discounted cumulative approach is the fact that it does not account for skew in visibility.
Counting items at discrete positions does not fully capture the fact that
minimal differences in relevance scores may translate into large differences in visibility for different groups  because of position bias that results in a large skew in the distribution of exposure.
For example in Figure \ref{fig:ranking}, the average utility of the 
items in $r_l$ that belong to the non-protected group is 0.35, whereas the average utility of the items that belong to the protected group is 0.33. This gives us  a difference of just 0.02. However, if we compute their exposure using DCG, (that is, if we discount utility logarithmically), the exposure for the items in the protected group is 25\% smaller than the  exposure for the items in  the non-protected group. 

The \textit{fairness of exposure} approach \cite{SJ18} generalizes the logarithmic discount, by assigning to
each position $j$ in the ranking a specific value that represents the importance
of the position, i.e., the fraction of users that examine an item at position $j$.
This is captured using a position discount vector $v$, where $v_j$ represents the importance of position $j$.
Note that we can get a logarithmic discount, if we set $v_j = \frac{1}{log_2(j+1)}$.
Rankings are seen as probabilistic. In particular, a  ranking of $N$ items in $N$ positions is modeled as a doubly stochastic $N \times N$  matrix $P$,  where $P_{i,j}$ is the probability that item $i$ is ranked at position $j$.

Given the position discount vector $v$, the exposure of item $i$ in ranking $P$ is defined as: 
\begin{equation}
Exposure(i|P) = \sum_{j=1}^{N}P_{i,j}v_j
\end{equation} 

The exposure of a group $G$ is defined as the average exposure of the items in the group:
\begin{equation}
Exposure(G|P) = \frac{1} {|G|}\sum_{i \in G} Exposure(i|P)
\end{equation}

In analogy to base rate statistical parity in classification, we get a \textit{demographic parity definition} of ranking fairness by asking that the two groups get the same exposure:
\begin{equation}
\frac{Exposure(G^+|P)}{Exposure(G^-|P)} = 1
\end{equation}

As with classification, we can also get additional statistical fairness definitions by taking into account the actual output, in this case, the utility of the items (e.g., their relevance to a search query $q$). This is called
\textit{disparate treatment constraint} in \cite{SJ18}, and it is expressed by asking that the exposure that the two groups receive is proportional to their average utility:
\begin{equation}
\frac{Exposure(G^+|P)}{Utility(G^+|q)} = \frac{Exposure(G^-|P)}{Utility(G^-|q)}
\end{equation}
%\textcolor{red}{what is q ??? in (6) and (7)}

Yet another definition, termed \textit{disparate impact}, considers instead of just the
exposure, the  impact of a
ranking. Impact is measured using the click-through rate (CTR), where CTR is estimated
as a function of both exposure and relevance. This definition asks that 
the impact of the ranking of the two groups is proportional to their average utility:
%\begin{footnotesize}
\begin{equation}
\frac{CTR(G^+|P)}{Utility(G^+|q)} = \frac{CTR(G^-|P)}{Utility(G^-|q)}
\end{equation}
%\end{footnotesize}

A fairness of exposure approach has also be taken to define individual fairness in rankings. Specifically, \textit{equity of attention}  \cite{BGW18} asks that
each item $i$ receives attention $a_i$ (i.e., views, clicks) that is proportional to its utility  $util_i$ (i.e., relevance to a given query):
%\begin{footnotesize}
\begin{equation}
\frac{a_1}{util_1} = \frac{a_2}{util_2}, \forall \,\, i_1, i_2
\end{equation}

In general, it is unlikely that equity of attention can  be satisfied in any single ranking. For example, multiple items may be similarly relevant for a given query, yet they obviously cannot all occupy the same ranking position. 
This is the case, for example with items with ID x329 and x23 in Figure \ref{fig:ranking}. 
To address this, \textit{amortized fairness} was proposed. 
 A sequence $\rho^1 \dots \rho^m$ of rankings offers \textit{amortized equity of attention} \cite{BGW18}, if each item receives cumulative attention proportional to its cumulative relevance, i.e.:

\begin{equation}\label{eq:equity}
\frac{\sum_{l=1}^{m} a^l_1}{\sum_{l=1}^{m}util^l_1} = \frac{\sum_{l=1}^{m}a^l_2}{\sum_{l=1}^{m}util^l_2}, \forall \,\, i_1, i_2.
\end{equation}

In this case, unfairness is defined as the distance between the attention and utility distributions.  
\begin{equation}\label{eq:unfair-weikum}
\begin{aligned}
unfairness(\rho^1,...,\rho^m) %&\frac{1}{N}\sum_{i=1}^n \left\|A_i - R_i\right\| \\
&= \sum_{i=1}^n \left\| \sum_{j=1}^m a_{i}^j - \sum_{j=1}^m util_{i}^j \right\|
\end{aligned}
\end{equation} 

A normalized version of this unfairness definition that considers the number $N$ of items  to be ranked and the number $m$ of rankings in the sequence, is proposed in \cite{DBLP:conf/medes/BorgesS19}. Formally: 
\begin{equation}
\begin{aligned}
unfairness(\rho^1,...,\rho^m) %&\frac{1}{N}\sum_{i=1}^n \left\|A_i - R_i\right\| \\
&= \frac{1}{N}\frac{1}{m}\sum_{i=1}^N \left\| \sum_{j=1}^m a_{i}^j - \sum_{j=1}^m r_{i}^j \right\|
\end{aligned}
\end{equation}

%The act of reducing unfairness imply reduction in the ranking quality. The original ranking is $\rho$, and the $NDCG-quality$ is calculated for a new item positioning ($\rho^*$) as: \\
%\begin{equation}
%\text{NDCG-quality@k}(\rho,\rho^*) = \frac{DCG@k(\rho^*)}{DCG@k(\rho)}.  
%\end{equation}

\subsection{Fairness in Recommenders} \label{sec:fair-recs}

%Typically, recommender systems trained toward accuracy. This way, the rankings produced by recommenders target at the main preferences of a user, while areas of lesser interest tend to be under-represented or even absent, leading many times to unbalanced recommendations with the risk of gradually narrowing down the user’s preferences. 
%Recommendation systems retrieve interesting items for users based on their profiles and their  history. Depending on the application and the recommendation system, history may include for example explicit user ratings of items, or, selection of items (e.g., views, clicks). 

In general, recommendation systems estimate a  score, or, rating,
$\hat{s} (u, i)$ for  a user  $u$ and an item $i$  that reflects the relevance of $i$ for $u$. Then, a recommendation list $I$ is formed for user $u$ that includes the  items having the highest estimated score for  $u$. %These scores can be seen as the utility scores in the case of recommenders. 
A simple approach to defining producer-side (that is, item-side) fairness for recommendations is to consider the recommendation problem as a classification problem where the positive class is the recommendation list. Then, any of the  fairness definitions in Section \ref{sec:levels} are readily applicable to defining producer-side fairness. 
Yet, another approach to defining producer-side fairness is to
consider the recommendation list $I$ as a ranked list $r$ and apply the various definitions described in Section \ref{sec:fair-rank}.

Next, we present a number of approaches proposed specifically for recommenders and discuss their relationship with approaches presented for ranking.

\textbf{Unfairness in predictions}. 
Recommendation systems have been widely applied
in several domains to suggest data items, like movies, jobs and courses. However, since predictions are based on observed data, they can inherit bias that may already exist. To handle this issue,  measures of consumer-side unfairness are introduced in \cite{DBLP:conf/nips/YaoH17}  that look into the discrepancy between the prediction behavior for \textit{protected} and \textit{non-protected} users.
Specifically, the proposed accuracy-based fairness metrics count the difference between the predicted and actual scores (i.e., the errors in prediction) of the data items recommended to users in the protected group $G^{+}$ and the items recommended to users in the non-protected group $G^{-}$. 
 
Let $N$ be the size of the recommendation list, 
$E_{G^+}[\hat{s}]_j$ and $E_{G^-}[\hat{s}]_j$ be the average predicted score ($\hat{s}$) that an item $j$ receives for the protected users and non-protected users respectively, and $E_{G^-}[\hat{s}]_j$ $E_{G^+}[s]_j$ and $E_{G^-}[s]_j$ be the corresponding average actual score ($s$) of item $j$.
Alternatives for defining unfairness can be summarized as follows: \\
%\begin{itemize}
%\item 
\textit{Value unfairness} ($U_{val}$) counts inconsistencies in estimation errors across groups, i.e., when one group  is given higher or lower predictions than their true preferences. That is: 
\begin{multline}
U_{val} = \frac{1}{N}\sum_{j=1}^n \big| \,(E_{G^+}[\hat{s}]_j  - E_{G^+}[s]_j) - \\ (E_{G^-}[\hat{s}]_j  - E_{G^-}[s]_j) \, \big|. 
\end{multline} 
Value unfairness occurs when one group of users is consistently given higher or lower predictions than their actual preferences.
For example, when considering course recommendations, value unfairness may suggest to male students engineering courses even when they are not interested in engineering topics, while female students not being recommended engineering courses even if they are interested in such topics. 

\textit{Absolute unfairness} ($U_{abs}$) counts inconsistencies in absolute estimation errors across user groups. That is: 
\begin{multline}
U_{val} = \frac{1}{n}\sum_{j=1}^N \big| \,|E_{G^+}[\hat{s}]_j  - E_{G^+}[s]_j| - \\ |E_{G^-}[\hat{s}]_j  - E_{G^-}[s]_j|\, \big|.
\end{multline} 
%\item 
Absolute unfairness is unsigned, so it captures a single statistic representing the quality of prediction
for each group. \textit{Underestimation unfairness} ($U_{under}$) counts inconsistencies in how much the predictions underestimate the true ratings. That is: 
\begin{multline}
U_{under} = \frac{1}{n}\sum_{j=1}^N \big| \,max\{0, E_{G^+}[s]_j - E_{G^+}[\hat{s}]_j\} - \\ max\{0, E_{G^-}[s]_j - E_{G^-}[\hat{s}]_j\} \, \big|. 
\end{multline} 
%\item 
Underestimation unfairness is important when missing recommendations are more critical than extra recommendations. 
For instance, underestimation may lead to top students not being recommended to explore topics they would excel in. 
\textit{Overestimation unfairness} ($U_{over}$) counts inconsistencies in how much the predictions overestimate the true ratings and is important when users may be overwhelmed by recommendations. That is: 
\begin{multline}
U_{over} = \frac{1}{n}\sum_{j=1}^N \big| \, max\{0, 
E_{G^+}[\hat{s}]_j - E_{G^+}[s]_j\} - \\ max\{0, 
E_{G^-}[\hat{s}]_j - E_{G^-}[s]_j\}, \big|. 
\end{multline} 
%\item 
Finally, \textit{non-parity unfairness} ($U_{par}$) counts the absolute difference between the overall average ratings of protected users and non-protected users. That is: 
\begin{equation}
U_{par} = \big| \, E_{G^+}[\hat{s}] - E_{G^-}[\hat{s}] \, \big|. 
\end{equation}
%\end{itemize} 

\textbf{Calibrated recommendations}. A calibration-based  approach to producer-side fairness  is proposed in \cite{DBLP:conf/recsys/Steck18}. 
A classification algorithm is considered to be well-calibrated if the predicted proportions
of the groups in the various classes agree with their actual proportions in the input data.
In analogy, the goal of a \textit{calibrated recommendation algorithm} is to reflect the interests of a user in the recommendations, and with their appropriate proportions. Intuitively, the proportion of the different groups of items in a recommendation list should be similar with their corresponding proportions in the history of the user. As an example, consider movies as the items to be recommended and  genre as the protected attribute.

For quantifying the degree of calibration of a list of recommended movies, with respect to the user’s history of played movies, this approach considers two distribution of the genre $z$ for each movie $i$, $p(z|i)$. Specifically, $p(z|u)$ is the distribution over genres $z$ of the set of movies in the history of the user $u$: 
\begin{equation}\label{eq:cal1}
p(z|u)=\frac{\sum_{i \in H}w_{u,i} \cdot p(z|i)}{\sum_{i \in H}w_{u,i}}, 
\end{equation} 
where $H$ is the set of movies played by user $u$ in the past and $w_{u,i}$ is the weight of movie $i$ reflecting how recently it was played by $u$. 

In turn, $q(z|u)$ is the distribution over genres z of the list of movies recommended to u: 
\begin{equation}\label{eq:cal2}
q(z|u)=\frac{\sum_{i \in I} w_{r(i)} \cdot p(z|i)}{\sum_{i \in I} w_{r(i)}}
\end{equation}
where $I$ is the set of recommended movies, and $w_r(i)$ is the weight of movie $i$ due to its rank $r(i)$ in the recommendation list. 

To compare these distributions, several methods can be used, like for example, the Kullback-Leibler (KL) divergence that is employed as a calibration metric. 
%\begin{footnotesize}
\begin{equation}\label{eq:kl}
C_{KL}(p,q) = \sum_z p(z|u)log(p(z|u)/\tilde{q}(z|u)),
\end{equation}
%\end{footnotesize}
where $p(z|u)$ is the target distribution and 
$\tilde{q}(z|u) = (1 - \alpha) q(z|u) - \alpha p(z|u)$ $\approx$ $q(z|u)$ with small $\alpha > 0$ is used to handle the fact that 
KL-divergence diverge for $q(z | u)$ = 0 and $p(z | u)$ $>$ 0.
 KL-divergence ensures that the genres that the user rarely played will also be reflected in the recommended list with their corresponding proportions; namely, it is sensitive to small discrepancies between distributions, it favors more uniform and less extreme distributions, and in the case of perfect calibration, its value is 0.

\textbf{Pairwise Fairness}.
Instead of looking at the scores that the items receive, \textit{pairwise fairness} looks at the relative position of pairs of items in a recommendation list.
The pairwise approach proposed in \cite{pairwise} is an accuracy-based one where the positive class includes 
the items that receive positive feedback from the user, 
such as clicks, high ratings, or increased user 
engagement (e.g., dwell-time).
For simplicity, in the following we will  assume only click-based feedback. 

Let $r(u, j)$
be 1 if user $u$ clicked on item $j$ and 0 otherwise.
Assume that  $\hat{r}(u, j)$ is the predicted probability that $u$ clicks on $j$ and $g$ a monotonic ranking function on $\hat{r}(u, j)$.
Let $I$ be the set of items, and $G^+$ and $G^-$  the group of protected and non-protected items, respectively.
\textit{Pairwise accuracy} is based on the probability
that a clicked item is ranked above another unclicked item, for the same user:
\begin{equation}
P(g(\hat{r}(u, j)) > g(\hat{r}(u, j')) \, | \, r(u, j) > r(u, j'), j, j' \in I)
\end{equation}
For succinctness, let $c_u(j, j') =  \mathbbm{1}[ g(\hat{r}(u, j)) > g(\hat{r}(u, j')] $.
The main idea is to ask that the two groups $G^+$ and
$G^-$ have similar  pairwise accuracy. Specifically, we achieve {\textit pairwise fairness} if:
 \begin{multline}
 P(c_u(j, j') | r(u, j) > r(u, j'), j \in G^+), j'  \in I = \\
   P(c_u(j, j') | r(u, j) > r(u, j'), j \in G^-, j'  \in I)
   \label{eq:pairwise}
 \end{multline}
This work also considers actual engagement by conditioning that the items have been engaged with the same amount.

We can also distinguish between intra- and inter-group fairness.  
\textit{Intra-group pairwise fairness} is achieved
if the likelihood of a clicked item being ranked above another
relevant unclicked item from the same group is the same independent
of the group, i.e, when both $j$ and $j'$ in Eq. \ref{eq:pairwise} belong to the same group, \textit{Inter-group pairwise fairness} is achieved if the likelihood of a clicked item being ranked above another
relevant unclicked item from the opposite group is the same independent
of the group, i.e., $j$ and $j'$ in Eq. \ref{eq:pairwise} belong to opposite groups.

\subsection{Fairness in Rank Aggregation}
\label{sec:group-rec-fairness-defs}
In addition to the efforts that focus on fairness in single rankings and recommendations, the problem of fairness in rank aggregation has also emerged. This problem arises when a number of ranked outputs is produced, and we need to aggregate these outputs in order to construct a new ranked consensus output. Typically, the problem of fair rank aggregation is largely unexplored. Only recently, some works study how to mitigate any bias introduced during the aggregation phase. This is done, mainly, under the umbrella of group recommendations, where instead of an individual user requesting recommendations from the system, the request is made by a group of users. 

As an example consider a group of friends that wants to watch a movie and each member in the group has his or her own likes and dislikes. The system needs to properly balance all users preferences, and offer to the group a list of movies that has a degree of relevance to each member. The typical way for doing so, is to apply a ranking or recommendation method to each member individually, and then aggregate the separate lists into one for the group \cite{DBLP:conf/er/NtoutsiSNK12,DBLP:journals/vldb/RoyACDY10}. For the aggregation phase, intuitively, for each movie, we can calculate the average score across all users in the group preference scores for the movie (\textit{average approach}). As an alternative, we can use the minimum function rather than the average one (\textit{least misery approach}), or even we can focus on how to ensure fairness by attempting to minimize the feeling of dissatisfaction within group members. 

Next, we present a fairness model for the general rank aggregation problem, and additional models defined for group recommendations.

%With the expansion of social media, another form of recommendations has emerged; namely group recommendations \cite{DBLP:conf/er/NtoutsiSNK12,DBLP:journals/vldb/RoyACDY10}. Instead of an individual user requesting recommendations from the system, a group can make a request as well. A standard example of group recommendations is the following: a group of friends wants to watch a movie and each one has their own likes and dislikes. The system needs to properly balance them, and offer to the group a list of items that has a degree of relevance to each member. The typical way for doing so, is to apply a recommendation method to each member individually, and then aggregate the separate lists into one for the group. For the aggregation phase, intuitively, for each item, we can calculate the average score across all users in the group preference scores for the item (\textit{average approach}). As an alternative, we can use the minimum function rather than the average one (\textit{least misery approach}). Several approaches recently focus on how to ensure fairness on group recommendations by attempting to minimize the feeling of dissatisfaction within group members. 

%

\textbf{Top-k Parity}. 
\cite{DBLP:journals/pvldb/KuhlmanR20} formalizes the fair rank aggregation problem as a constrained optimization problem. Specifically, given a set of rankings, the fair rank aggregation problem returns the closest ranking to the given set of rankings that satisfies a particular fairness criterion. 

Given that each data item has a protected attribute that partitions the dataset in $m$, $m \geq  2$, disjoint groups $G = \{G_1, \ldots, G_m\}$, this work uses as a fairness criterion a general formulation of statistical parity for rankings that considers the top-k prefix of the rankings, namely the \textit{top-k parity}. Formally, given a ranking of $n$ data items belonging to mutually exclusive groups $G_i \in G$, and $0 \leq k \leq n$, the ranking satisfies \textit{top-k parity} if the following condition is met for all, $G_i, G_j \in G$, $i \neq j$: 
\begin{equation}
P(\rho(x) \leq k | x \in G_i) = P(\rho(x) \leq k | x \in G_j), 
\end{equation}
where $\rho(x)$ denotes the position of the data item $x$ in the ranking.

\textbf{Dissatisfaction fairness}. For counting fairness, a measure of quantifying the satisfaction, or \textit{utility}, of each user in a group given a list of recommendations for this group, can be used, namely by checking how relevant the recommended items are to each user \cite{DBLP:conf/recsys/LinZZGLM17}.
Formally, given a user $u$ in a group $G$ and a set $I$ of $N$ items recommended to $G$, the individual utility $util(u,I)$ of the items $I$ for $u$ can be defined as the average items utility, normalized in $[0,1]$, with respect to their relevance for $u$, $rel(u,i)$:
\begin{equation}
util(u,I) = \frac{\sum_{i\in I}rel(u,i)}{N \times rel_{max}}
\end{equation}
%, or 
%\begin{equation}
%U(u,I) = \frac{\sum_{i\in I}rel(u,i)}{\sum_{i\in I(u,N)}rel(u,i)}
%\end{equation}
where 
%$I(u,N)$ denotes the set of items, which are among the top-$N$ favourite items of $u$, and 
$rel_{max}$ denotes the maximum value $rel(u,i)$ can take. 
In turn, the overall satisfaction of users about the group recommendation quality, or \textit{group utility}, is estimated via aggregating all the individual utilities. This is called \textit{social welfare}, $SW(G,I)$, and is defined as: 
\begin{equation}\label{eq:sw}
SW(G,I) = \frac{\sum_{u\in G}util(u,I)}{|G|}.
\end{equation}

Then, for estimating fairness, we need to compare the utilities of the users in the group. Intuitively, for example, a list that minimizes the dissatisfaction of any user in the group can be considered as the most fair. In this sense, fairness enforces the least misery principle among users utilities, emphasising the gap between the least and highest utilities of the group members. Following this concept, fairness can be defined as: 
\begin{equation}\label{eq:fsw1}
F(G,I)= min\{util(u,I), \forall u \in G \}.
\end{equation}

Similarly, fairness can encourage the group members to achieve close utilities between each other using variance: 
\begin{equation}\label{eq:fsw2}
F(G,I)= 1 - Variance(\{util(u,I), \forall u \in G \}).
\end{equation}

\textbf{Pareto optimal fairness}. 
Instead of computing users' individual utility for a list of recommendations  by summing up the relevance scores of all items in the list  \cite{DBLP:conf/recsys/LinZZGLM17}, the item positions in the recommendation list can be considered \cite{DBLP:conf/sac/Sacharidis19}.  Specifically, the solution for making fair group recommendations is based on the notion of Pareto optimality, which means that an item $i$ is \textit{Pareto optimal} for a group if there exists no other item $j$ that ranks higher according to all users in the group, i.e., there is no item $j$ that dominates item $i$. \textit{N-level Pareto optimal}, in turn, is a direct extension that contains items dominated by at most $N-1$ other items, and is used for identifying the $N$ best items to recommend. Such a set of items is fair by definition, since it contains the top choices for each user in the group.

\textbf{Fairness in package-to-group recommendations.} 
Given a group $G$, an approach to fair package-to-group recommendations is to recommend to $G$ a package of items $P$, by requiring that for each user $u$ in $G$, at least one item high in $u$'s preferences is included in $P$ \cite{DBLP:conf/www/SerbosQMPT17}.
Even if such a resulting package is not the best overall, it is fair, since there exists at least one item in $P$ that satisfies
each user in $G$.

Specifically, two different aspects of fairness are examined \cite{DBLP:conf/www/SerbosQMPT17}: 
(a) fairness proportionality, ensuring that each user finds a sufficient number of items in the package that he/she likes compared to items not in the package, and (b) fairness envy-freeness, ensuring that for each user there is a sufficient number of items in the package that he/she likes more than other users do. Formally: 

 \textit{m-proportionality.} For a group $G$ of users and a package $P$, the $m$-proportionality of $P$ for $G$ is defined as: 
%\begin{footnotesize}
\begin{equation}\label{eq:fprop}
F_{prop}(G,P) = \frac{|G_P|}{|G|}, 
\end{equation} 
%\end{footnotesize}
where $G_p \subseteq G$ is the set of users in $G$ for which $P$ is $m$-proportional. In turn, $P$ is $m$-proportional to a user $u$, if there exist at least $m$ items in $P$, such that, each one is ranked in the top-$\delta$\% of the preferences of $u$ over all items in $I$, for an input parameter $\delta$.

 \textit{m-envy-freeness}. For a group of users $G$ and a package $P$, the $m$-envy-freeness of $P$ for $G$ is defined as: 
%\begin{footnotesize}
\begin{equation}\label{eq:fenvy}
F_{ef}(G,P) = \frac{|G_{ef}|}{|G|}, 
\end{equation}
%\end{footnotesize}
where $G_{ef} \subseteq G$ is the set of users in $G$ for which $P$  is $m$-envy-free. 
In turn, $P$ is m-envy-free for a user $u$, if $u$ is envy-free for at least $m$ items in $P$, i.e., each item's $rel(u,i)$ is in the top-$\delta$\% of the preferences in the set $\{rel(v,i) : v \in G\}$.

\textbf{Sequential hybrid aggregation}. 
An approach for fair sequential group recommenders targets two independent objectives \cite{DBLP:conf/sac/StratigiNPS20}. 
The first one considers the group as an entity and aims at offering the best possible results, by maximizing the overall group satisfaction over a sequence of recommendations. 
The satisfaction of each user $u_i$ in a group $G$ for the group recommendation $Gr_j$ received at the $j$-$th$ round of recommendations, is computed by comparing the quality of recommendations that $u_i$ receives as a member of the group over the quality of recommendations $u_i$ would have received as an individual. 
Given the list $A_{u_i,j}$ with the top-$N$ items for $u_i$, the user's satisfaction is calculated based on the group recommendation list, i.e., for every item in $Gr_j$, we sum the utility scores as they appear in each user's $A_{u_i,j}$, over the ideal case for the user, by sum the utility scores of the top-$N$ items in $A_{u_i,j}$. Formally: 
\begin{equation} \label{eq:sequentialsatisfaction}
    sat(u_i,Gr_j)=\frac{\sum_{d_z \in Gr_j}util_j(u_i,d_z)}{\sum_{d_z \in A_{u_{i},j}}util_j(u_i,d_z)}
\end{equation} 

The second objective considers the group members independently and aims to behave as fairly as possible towards all members, by minimizing the variance between the user satisfaction scores. Intuitively, this variance represents the potential disagreement between the users in the group. Formally, the disagreement is defined as: 
\begin{align}
    & groupDis(G,\mathcal{GR})=\nonumber\\
    & max_{u_i \in G}satO(u_i, \mathcal{GR}) - min_{u_i\in G}satO(u_i, \mathcal{GR}) \label{eq:maxmin}
\end{align}
where $satO(u_i, \mathcal{GR})$ is the overall satisfaction of $u_i$ for a sequence $\mathcal{GR}$ of recommendations defined as the average of the satisfaction scores after each round. That is, group disagreement is the difference in the overall satisfaction scores between the most satisfied and the least satisfied user in the group. When this measure takes low values, the group members are all satisfied to the same degree.

\begin{table*}[t] 
\caption{Fairness definitions taxonomy in Rankings, Recommenders and Rank Aggregation.}
%\caption{Fairness definitions taxonomy with respect to Individual (IF) and Group (GF) Fairness, Consumer (CF) and Producer (PF) Fairness, and Fairness for Single (FS) and Sequential (FSe) interactions, in Rankings, Recommenders and Group-Recommenders.} 
\begin{scriptsize}
\begin{tabular}{|l||c|c||c|c||c|c||l|}
\hline
 & \rotatebox{90}{\textbf{Individual~}} & \rotatebox{90}{\textbf{Group}} & \rotatebox{90}{\textbf{Consumer}} & \rotatebox{90}{\textbf{Producer}} & \rotatebox{90}{\textbf{Single}} & \rotatebox{90}{\textbf{Multiple}} & \rotatebox{90}{\textbf{Criterion}} \\ \hline \hline
\textit{\textbf{Rankings}}  &             &             &             &             &             &             &              \\ \hline
Fairness constraints \cite{CSV18} & & \checkmark &             & \checkmark & \checkmark &              & position \\ \hline
Discounted cumulative fairness \cite{YS17} & & \checkmark &             & \checkmark & \checkmark &              & position \\ \hline
Fairness of exposure \cite{SJ18} & & \checkmark &             & \checkmark & \checkmark &              & position/utility \\ \hline
Equity of attention \cite{BGW18} & \checkmark &             &             & \checkmark & & \checkmark              & position/utility \\ \hline \hline
\textit{\textbf{Recommenders}} & & & & & & & \\ \hline
Calibrated recommendations \cite{DBLP:conf/recsys/Steck18} & & \checkmark & & \checkmark & \checkmark &               &  number of items
 \\ \hline
Value/Absolute unfairness \cite{DBLP:conf/nips/YaoH17} & & \checkmark & \checkmark &             & \checkmark &  & error on predictions\\ \hline
Under/Overestimation unfairness \cite{DBLP:conf/nips/YaoH17} & & \checkmark & \checkmark &             & \checkmark & & error in ratings\\ \hline
Non-parity unfairness \cite{DBLP:conf/nips/YaoH17} & & \checkmark & \checkmark &             & \checkmark & & ratings \\ \hline \hline
\textit{\textbf{Rank Aggragation}} & & & & & &  &\\ \hline 
Top-k parity \cite{DBLP:journals/pvldb/KuhlmanR20} & & \checkmark & \checkmark &             & \checkmark &  & position
\\ \hline 
Dissatisfaction fairness \cite{DBLP:conf/recsys/LinZZGLM17} & \checkmark & & \checkmark &             & \checkmark &  & user satisfaction
\\ \hline
Pareto optimal fairness \cite{DBLP:conf/sac/Sacharidis19} & \checkmark & & \checkmark &             & \checkmark &  & position\\ \hline
Proportionality fairness \cite{DBLP:conf/www/SerbosQMPT17} & \checkmark & & \checkmark &             & \checkmark &   &  number of items\\ \hline
Envy-freeness fairness \cite{DBLP:conf/www/SerbosQMPT17} & \checkmark & & \checkmark &             & \checkmark &  & number of items\\ \hline 
Sequential hybrid aggregation \cite{DBLP:conf/sac/StratigiNPS20} & \checkmark & & \checkmark &             &             & \checkmark &   user satisfaction\\ \hline
\end{tabular} 
\end{scriptsize}
\label{tab:fairdefs} 
\end{table*}

%of Fairness Definitions in Rankings and Recommender
\subsection{Summary} 
In short, we can categorize the various definitions of fairness used in rankings, recommendations and rank aggregation methods based on the \textit{Level} and \textit{Side} of fairness, and their \textit{Output Multiplicity}. Specifically, regarding \textit{Level}, fairness can be distinguished between individual and group fairness. 
%In turn, group fairness can fall into three categories, namely, \textit{independence} fairness, when the outcome of a model is independent from the protected attributes, \textit{separation} fairness, when the outcome of a model is independent from the protected attributes conditional on a target variable, and \textit{sufficiency} fairness, when the true outcome of a model and the protected attributes are independent \cite{barocas-hardt-narayanan}. 
The \textit{Side} dimension considers consumer and producer fairness, while the \textit{Output Multiplicity} dimension considers single and multiple outputs. 

Table \ref{tab:fairdefs} presents a summary of the various definitions of fairness. We make several observations. Specifically, we observe that these fairness definitions are based on one of the following criteria: (a) position of item in the ranking or recommendation list (b) item utility (c) prediction error (d) rating, and (e) number of items. User satisfaction is defined through item utility. 
In rankings, fairness is typically defined for the items to be ranked, matching all existing definitions to producer fairness. Except equity of attention, the definitions are group-based, and to position bias, they use constraints on the proportion of items in the top positions, or are based on the idea of providing exposure proportional to utility. 
In recommenders, all definitions are group-based. In this case, the distinction between consumer fairness and producer fairness makes more sense, given that they focus either on the individuals that receive a recommendation or the individuals that are recommended. Nevertheless, most existing works target consumer fairness. 
%even if both categories fit well with recommenders, 

%this example is used earlier:  For instance, when considering consumer fairness, we may ask that race does not influence the job recommendations that individuals receive, while for producer fairness, we may ask that the political orientation of an article does not affect the recommendation result.

In rank aggregation, only recently, and mainly for group recommenders, fairness is considered when we need to aggregate a number of ranked outputs in order to produce a new ranked consensus output. Specifically, for group recommenders, the goal is to evaluate if the system takes into consideration the individual preferences of each single user in the group, making all approaches in the research literature to target at individual and consumer fairness. 
Only recently, there are few approaches that focus on another form of fairness that is applicable when we consider a sequence of rankings, or recommendations, instead of just a single one. 

In what follows, we will study how these models and definitions of fairness are applied to algorithms.

%\textcolor{red}{Interesting questions to discuss:}
%2. Are these definitions bound to specific algorithm? For example, classification, matrix factorization? If yes, can they be generalized to other cases?
%3. We have seen models for rankings and models for recommendations. The question that arises is whether a model of fairness in rankings can be used for recommendations and vice versa. 

\section{Methods for Achieving Fairness} \label{sec:framework}

Taking a cross-type view, we present in this section a taxonomy to organize and place related works into perspective. Specifically, while in Figure \ref{fiq:fairmethods}(a) we show the traditional way that results in ranked outputs, in Figure \ref{fiq:fairmethods}(b), we present the various options and the general distinction of the methods for generating fair ranked outputs and recommendations. 
Namely, these methods are distinguished into the following categories: 

\begin{figure*}
\centering
ai\includegraphics[width=1.8\columnwidth]{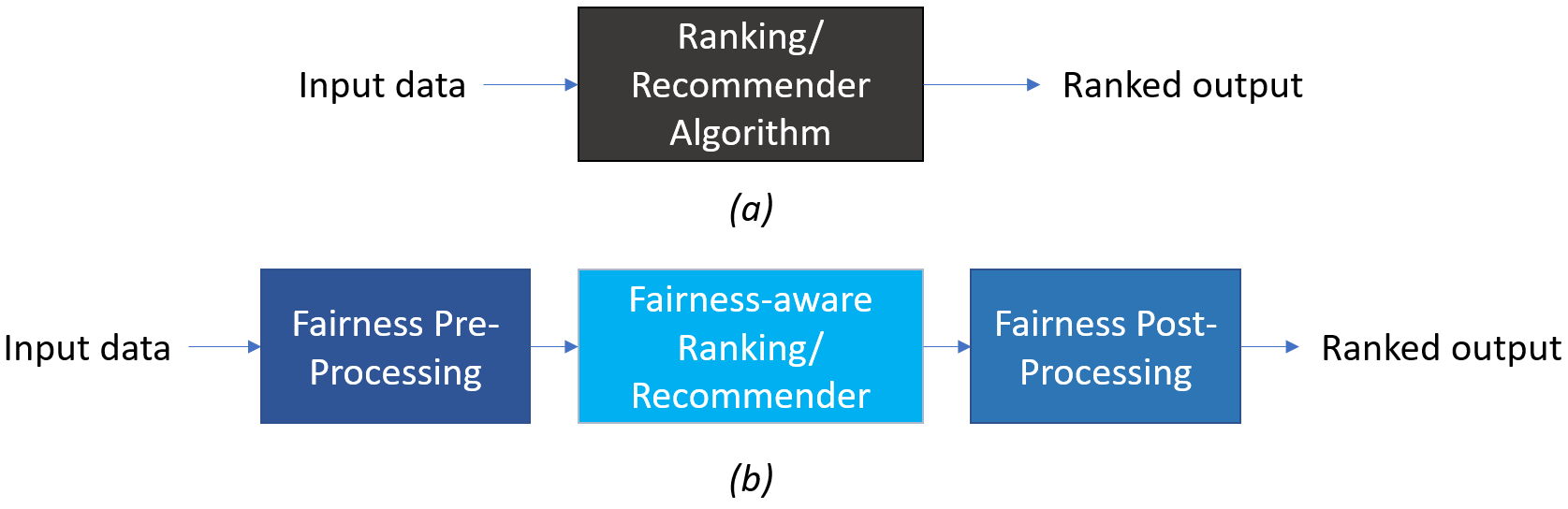}
\caption{The general distinction of the methods for ensuring fair ranked outputs.}
\label{fiq:fairmethods}
\end{figure*}

\begin{itemize}
    \item \emph{Pre-processing methods} aim at transforming the data to remove any underlying bias or discrimination. Typically, such methods are application agnostic, and consider bias in the \textit{training data} (which they try to mitigate). Bias in the data may be produced due to the data collection process (for example, based on decisions about the pieces of data we collect or not, or what assumptions we make for the missing values), or even when using data in a different way than intended during collection.

    \item \emph{In-processing methods} aim at modifying existing or introducing new algorithms that result in fair rankings and recommendations, e.g., by removing bias and discrimination during the model training process. Typically, such methods targets at learning a model with no bias, while considering fairness during the training of a model, for example, by incorporating changes into the objective function of an algorithm by a fairness term or imposing fairness constraints, without offering any guarantees about fairness on the ranked outputs. 
    
    \item \emph{Post-processing methods} modify the output of the algorithm. Typically, such methods can only treat the ranking or recommendation algorithm as a black box without any ability to modify it, and to improve fairness they re-rank the data items of the output. Naturally, in post-processing methods, fairness comes at the cost of accuracy, since by definition the methods transform the optimal output. On the other hand, a clear advantage of the post-processing methods is that they offer ranked outputs that are easy to understand, when comparing their outputs with the outputs before any application of a post-processing fairness method. 
\end{itemize}

Next, we will use this taxonomy to organize and present the related works that we describe in the following sections.

\section{Pre-processing Methods} \label{sec:pre-proc}

Bias in the underlying data on which systems are trained can take two forms. 
\emph{Bias in the rows} of the data exists when there are not enough representative individuals from minority groups. For example, according to a Reuters article \cite{amazon-jobs-2018}, Amazon's experimental automated system to review job applicants' resumes showed a significant gender bias towards male candidates over females that was due to historical discrimination in the training data.

\emph{Bias in the columns} is when features are biased (correlated) with sensitive attributes. For example, zip code tends to predict race due to a history of segregation \cite{amazon-race-2016}. Direct discrimination occurs when protected attributes are used explicitly in making decisions (i.e., \emph{disparate treatment}). More pervasive nowadays is indirect discrimination, in which protected attributes are not used but reliance on variables correlated with them leads to significantly different outcomes for different groups, also known as \emph{disparate impact}.

To address bias and avoid discrimination, several methods have been proposed for pre-processing data. Many of these methods are studied in the context of classification,  while a few have been proposed in the context of recommender systems. 

\subsection{Suppression}
To tackle bias in the data, a na\"{\i}ve approach used in practice is to simply \emph{omit the protected attribute} (say, race or gender) when training the classifier \cite{DBLP:journals/kais/KamiranC11}.

Simply excluding a protected variable is insufficient to avoid discriminatory predictions, as any included variables that are correlated with the protected variables still contain information about the protected characteristic, and the classifier still learns the discrimination reflected in the training data. For example, answers to personality tests identify people with disabilities \cite{wsj-2014}.
Word embeddings trained on Google News articles exhibit female/male gender stereotypes  \cite{DBLP:conf/nips/BolukbasiCZSK16}.
To tackle such dependencies, one can further find the attributes that correlate most with the sensitive attribute and remove these as well.

\subsection{Class Relabeling}

This approach, also known as massaging \cite{DBLP:journals/kais/KamiranC11}, changes the labels of some objects in the dataset in order
to remove the discrimination from the input data. A good selection of which labels to
change is essential. The idea is to consider a subset of data from the minority group as promotion candidates, and change their class label. Similarly, a subset of the majority group is chosen as demotion candidates.
To select the best candidates for relabeling, a ranker is used that ranks the objects based on their probability of having
positive labels. For example,
 a Na\"{\i}ve Bayesian classifier can be used for both  ranking and learning \cite{4909197,DBLP:journals/kais/KamiranC11}.
Then, the top-k minority, for promotion, objects  and the bottom-k  majority, for
demotion, objects are chosen.
The number $k$ of pairs needed to be modified to make a dataset $D$ discrimination-free
can be calculated as follows.

\begin{figure}
\centering
\includegraphics[width=0.9\columnwidth]{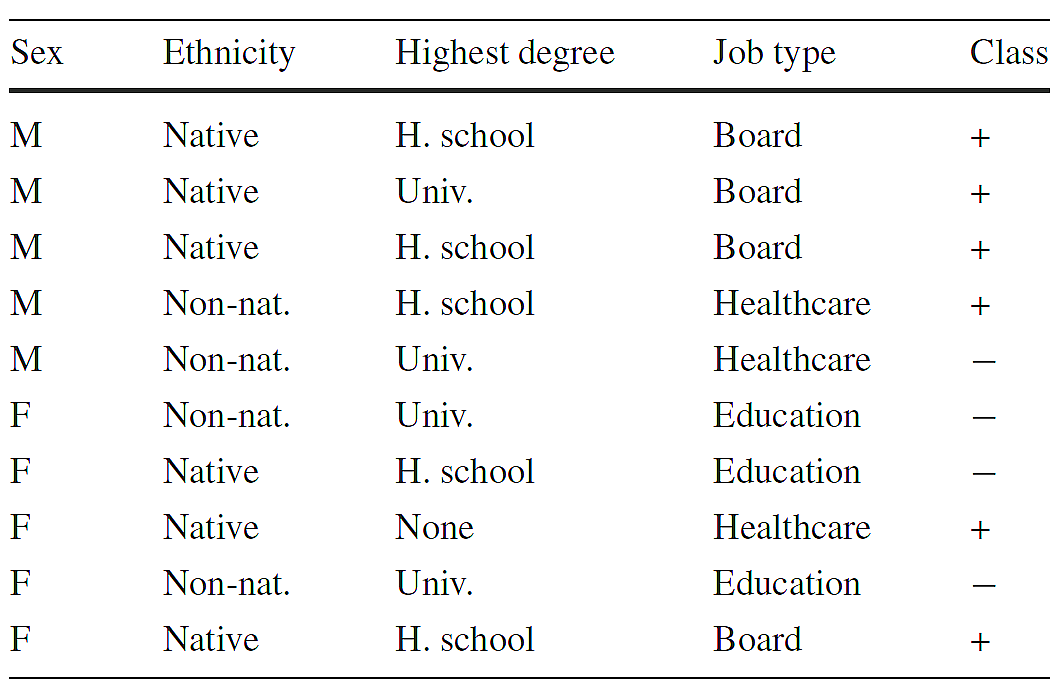}
\caption{Job application example \cite{DBLP:journals/kais/KamiranC11}.}
\label{fig:jobapp}
\end{figure}

Let us assume as before two groups, namely,  the protected group $G^+$ and the non-protected (or, privileged) group $G^-$.
If we modify $k$  objects from each group, the resulting discrimination will be:
\begin{align}
 & \frac{p^- - k}{|G^-|} - \frac{p^+ + k}{|G^+|} = disc(D) - k (\frac{1}{|G^+|} + \frac{1}{|G^-|}) = \nonumber\\
 &  disc(D) - ( k \frac{|D|}{|G^-||G^+|} )\label{eq:relabel}
\end{align}

%\begin{align}
% & \frac{p_w - k}{|D_w|} - \frac{p_w + k}{|D_b|} = disc(D) - k (\frac{1}{|D_b|} + \frac{1}{|D_w|}) = \nonumber\\
% &  disc(D) - ( k \frac{|D|}{|D_w||D_p|} )\label{eq:relabel}
%\end{align}

To reach zero discrimination, the number of modifications needed is:
\begin{equation}
k = \frac{disc(D)  \times |G^-| \times |G^+|}{|D|}
\end{equation}
 where $p^+$ ($p^-$) are the number of positive objects that belong to the minority group (majority group). Discrimination $disc(D)$ in $D$ is the probability of being in the positive class between the objects in the minority group versus those in the majority group.

\begin{figure}
\centering
\includegraphics[width=\columnwidth]{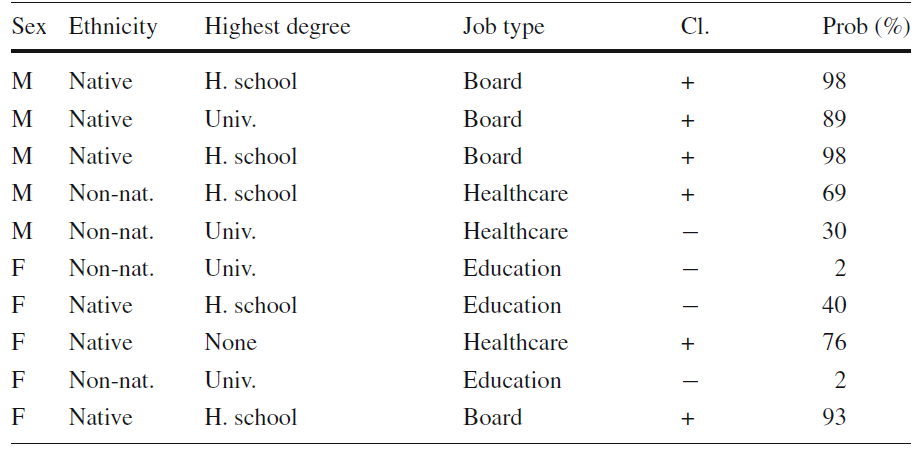}
\caption{Job applications with positive class probability}
\label{tab:relabeling}
\end{figure}

\begin{figure}
\centering
\includegraphics[width=\columnwidth]{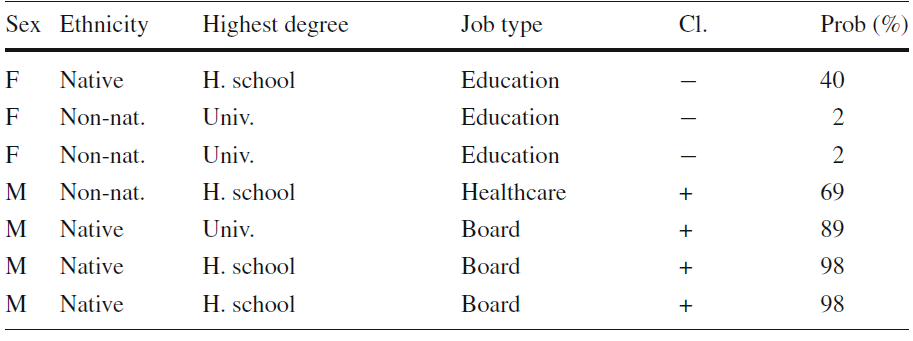}
\caption{Promotion and demotion candidates}
\label{tab:promotion}
\end{figure}

For example, consider the dataset in Figure \ref{fig:jobapp}.  This dataset contains the Sex, Ethnicity, and Highest Degree for 10 job applicants, the Job Type they applied for, and the outcome of the selection procedure, Class. We want to learn a classifier to predict the class of objects for which the predictions are non-discriminatory toward females. We can rank the objects by their positive class probability given by a  a Na\"{\i}ve Bayes classification model. Figure \ref{tab:relabeling} shows an extra column that gives the probability that each applicant belongs to the positive class.
In the second step, we arrange the data separately for female applicants with class $-$ in descending order and for male applicants with class $+$ in ascending order with respect to their positive class probability. The ordered promotion and demotion candidates are given in Figure \ref{tab:promotion}.
To reach zero discrimination, the number of modifications needed is: 
\begin{equation}
k = \frac{disc(D)  \times |G^{F}-| \times |G^{M}+|}{|D|} = \frac{0.4 \times 5 \times 5}{10} =1 
\end{equation}

We relabel the highest scoring female with a negative label and the lowest scoring male with a positive label. Then, the discrimination becomes zero. The resulting dataset will be used for training a classifier.

The problem of classification without discrimination w.r.t. a sensitive attribute is 
a multi-objective optimization problem. Lowering the discrimination will result in lowering the accuracy and vice versa.
%: we can trade in accuracy in order to reduce the discrimination.

% on the one hand, the more discrimination we allow for, the higher accuracy we can obtain while on the other hand,

\subsection{Reweighing}
The previous approach is rather intrusive as it changes the labels of the objects.
Instead of that, weights can be assigned to the objects  to compensate for the bias \cite{DBLP:journals/kais/KamiranC11}. 
The idea is to assign lower weights to objects that have been deprived or favored.
Then, the weights can be used directly in any method based on frequency counts.

%By carefully choosing the weights, the training dataset can be
%made discrimination-free w.r.t. the sensitive attribute without having to change any of the labels. In a nutshell, to compensate for the bias, we will assign lower weights to objects that have been deprived or favored.
%The weights on the tuples can be used directly in any method based on frequency counts.
%Clearly there will be a trade-off between the accuracy and the discrimination of the classifier.

A frequently used family of analytical methods  are grouped under \emph{propensity score matching} \cite{JSSv042i08}. Such methods model the probability of each object or group receiving the treatment and use these predicted probabilities or ``propensities'' to make up for the confounding of the treatment with the other variables of interest and balance the data.

A simple \emph{probability-based reweighing} method is the following \cite{DBLP:journals/kais/KamiranC11}.
Let us consider the sensitive attribute $S$. Then, every object $x$ will be assigned a  weight:
\begin{equation}\label{eq:rewiegh}
w(x) = \frac{P_{exp}(S=x(S) \wedge class = x(class))}{P_{obs}(S=x(S) \wedge class = x(class))}
\end{equation}
i.e., the weight of an object will be the expected probability to see an instance with its sensitive
attribute value and class given independence, divided by its observed probability.

For example, consider the dataset in Figure \ref{fig:jobapp}. If the dataset is unbiased, then the sensitive attribute $S$ (i.e., sex in our example) and the class are statistically independent. Then, the expected probability for females to be promoted would be:  $P_{exp}(sex=F \wedge class = +) = 0.5 * 0.6 = 0.3$. In reality, however, the observed probability  based on the dataset is $P(sex=F \wedge class = +) = 0.2$. Hence, one can use a re-weighting factor $w(x)=0.3/0.2=1.5$ to balance the bias in the dataset.

\emph{Entropy balancing} aims at covariate balance in data for binary classification \cite{hainmueller2012}. It relies on a maximum entropy reweighting scheme that calibrates individual weights so that the  reweighed groups satisfy a set of balance constraints
that are imposed on the sample moments of the covariate distributions. The balance constraints
ensure that the reweighed groups match exactly on the specified moments adjusting in this way inequalities in representation. The generated weights can be passed to any standard classifier.  
%Entropy balancing thereby adjusts inequalities in representation with respect to the first, second, and possibly higher moments of the covariate distributions.
%This reduces model dependence with regression or other standard estimators in the preprocessed data.

%This reduces model dependence for the subsequent estimation of treatment effects with regression or other standard estimators in the preprocessed data.
%These balance improvements can reduce model dependence for the subsequent estimation of treatment effects.

\emph{Adaptive Sensitive Reweighing}   uses a convex model to estimate distributions of underlying labels with which to adapt weights \cite{Krasanakis2018}. It assumes that there exists an (unobservable)
underlying set of class labels corresponding to training samples that, if predicted, would yield unbiased classification with
respect to a fairness objective. It searches for sample weights that make weighted training on the original dataset also train towards those labels, without explicitly observing them.

More specifically, consider a binary probabilistic classifier, which produces probability estimates $\hat{P}(Y = y_i ) = 1 - \hat{P}(Y \neq y_i )$.
For training samples $i$ with features $x_i$ and class labels $y_i$, there exists an underlying (i.e. unobservable) class
labels $\tilde{y_i}$ that yield estimated labels $\hat{y_i}$ which conform to designated
fairness and accuracy trade-offs. The training goal is to minimize  both weighted error on observed labels as well as the distance between weighted observed labels and unweighed underlying labels:
\begin{equation}
min\sum_i{w_i\hat{P}(\hat{y_i} \neq y_i)}
\end{equation}
\begin{equation}
min\sum_i({w_i\hat{P}(\hat{y_i} \neq y_i) - \hat{P}(\hat{y_i} \neq \tilde{y_i})})^2
\end{equation}

To simultaneously adjust training weights alongside classifier training, a classifier-agnostic iterative approach is proposed: first, a classifier is fully trained based on uniform weights, and
then, the method appropriately readjusts those weights. This process is repeated until convergence.

%This  method can be applied on multiple types of fairness objectives, such as trade-offs between accuracy and disparate impact elimination or disparate mistreatment elimination.

%\cite{Zafar2017}

%Existing data preprocessing techniques, being suppression of the sensitive attribute, massaging the dataset by changing class labels, and reweighing or resampling the data to remove discrimination without relabeling instances.

\subsection{Data Transformation}
A common theme is the importance of balancing discrimination control against utility of the processed data.
This can be formulated as an optimization problem for producing preprocessing
transformations that trade off discrimination control, data utility, and individual distortion \cite{DBLP:conf/nips/CalmonWVRV17}.
Assuming $S$ is the one or more protected (sensitive) variables, $X$  denotes other non-protected variables, and $Y$
is an outcome random variable. The goal is to determine a randomized mapping $P_{\hat{X,Y}|X,Y,S}$ that transforms both the training data and the test data. The mapping should satisfy three properties.

$-$ \emph{Discrimination Control}. The first objective is to limit the dependence of the transformed outcome
$\hat{Y}$ on the protected variables $S$, which requires the conditional distribution $P_{\hat{Y}|S}$ to be close to a target distribution $P_{Y_T}$ for all values of $S$.

$-$  \emph{Distortion Control}. The mapping $P_{\hat{X,Y}|X,Y,S}$  should satisfy distortion constraints to reduce or avoid certain large changes (e.g. a very low credit score being mapped to a very high credit score).

$-$  \emph{Utility Preservation}. The distribution of ($\hat{X}, \hat{Y}$) should  be statistically close to the distribution of $(X, Y)$. This is to ensure that a model learned from the transformed data (when averaged over the protected variables $S$) is not
too different from one learned from the original data. For example, a bank' s existing policy for approving loans does not change much when learnt over the transformed data.

\subsection{Database Repair}
Handling bias in the data can be considered a \emph{database repair problem}.
One approach is to \emph{remove information about
the protected variables from the set of covariates} to be used in predictive models \cite{FFM+15,DBLP:journals/corr/LumJ16}.
A test for disparate impact based on how well the protected class can be predicted from the other attributes and a data repair algorithm for numerical attributes have been proposed \cite{FFM+15}. The algorithm ``strongly preserves rank'', which means it changes the data in such a way that predicting the class is still possible. %  continues choosing stronger (higher ranked) applicants over weaker ones.
A chain of conditional models can be used for both protecting and adjusting variables of arbitrary type \cite{DBLP:journals/corr/LumJ16}. This framework allows for an arbitrary number of variables to be adjusted and for each of these variables and the protected variables to be continuous or discrete.

Another data repair approach is based on measuring the discriminatory causal influence of the protected attribute on the outcome of an algorithm. This approach removes discrimination by repairing the training data in order to \emph{remove
the effect of any discriminatory causal relationship between the protected attribute and classifier predictions} \cite{DBLP:conf/sigmod/SalimiRHS19}.
%, without assuming adherence to an underlying causal models.
This work introduced the notion of  \emph{interventional fairness}, which ensures that the protected
attribute does not affect the output of the algorithm in any configuration of the system obtained by
fixing other variables at some arbitrary values. 
The system repairs the input data by inserting or removing tuples, changing the empirical probability distribution
to remove the influence of the protected attribute on the outcome through any causal pathway that includes inadmissible attributes, i.e. attributes that should not influence the protected attribute.
%This notion correctly captures group fairness and is the strongest notion of fairness that is testable from data.
%\cite{Hajian2013} considers a similar approach for 

%\cite{DBLP:conf/nips/CalmonWVRV17,ZWS+13}.

% Prior work neither presents general and principled optimization frameworks for trading off accuracy and fairness, nor allows connections to be made to the broader statistical learning and information theory literature via probabilistic descriptions.

%The goal is to find an intermediate representation of the data \cite{ZWS+13}  that preserves as much information about the individual's attributes as possible, while simultaneously obfuscates aspects of
%it, removing any information about membership with respect to the protected group. The intermediate representation should encode the data as well as possible, and it  is sanitized in the sense that it should be blind to whether or not
%the individual is from the protected group.
%The main idea  is to map each individual, represented as a data point in a given input space, to a probability distribution in a new representation space. The aim of this new representation is to lose any information that can identify whether the person belongs to
%the protected subgroup, while retaining as much other information as possible.

%However, this prior work neither presents general and principled optimization frameworks for trading off these two criteria, nor allows connections to be made to the broader statistical learning and information theory literature via probabilistic descriptions.

\subsection{Data Augmentation}
A different approach is augment the training data with additional data \cite{DBLP:conf/wsdm/RastegarpanahGC19}. This framework  starts from an existing matrix factorization
recommender system that has already been trained with some input (ratings) data, and adds new users who provide ratings of existing items. The new users' ratings,  called antidote data, are chosen
 so as to improve a socially relevant property of the recommendations that are provided to the original users. The proposed framework includes measures of both individual and group unfairness.

\subsection{Summary of Pre-processing Methods}

Table \ref{tab:Preprocessing} summarizes pre-processing approaches to fairness based on whether they focus on bias in rows or columns, the level of fairness (individual or group) and the algorithm that will use the pre-processed data.

Many of the  pre-processing methods are studied in the context of classification/ranking. Suppression is a simple, brute-force approach that does not depend on the algorithm. On the downside, the algorithm may still learn the discrimination from correlated attributes. Trying to remove these attributes as well can seriously hurt the value of the dataset.

The class relabeling approach works with different rankers (e.g., a Naive Bayes classifier, or  nearest neighbor classifier). However, its objective is to lower the discrimination, which will result in lowering the accuracy and vice versa. Finding the right balance is challenging. Moreover, it is intrusive as it changes the dataset.

The reweighing methods are parameter-free as they do not rely on a ranker. Hence, they can work with any ranking algorithm as long as they leverage frequencies.
Adaptive Sensitive Reweighing has an additional classification overhead since it simultaneously adjusts training weights alongside classifier training through an iterative approach which is repeated until convergence.
%There is also a large literature on reweighting methods in economics and statistics that could be worth looking at in the context of ranking and recommendation methods.

Data transformation methods work with different classifiers but they can be applied only on numerical datasets and they modify the data.

The aforementioned approaches modify the training data,  explicitly (e.g., by suppressing attributes or changing class labels) or implicitly, e.g., by adding weights. 
Data augmentation leaves the training data as is and just augment it with additional data. 
 One data augmentation method has been proposed in the context of recommender system, and in particular for matrix factorization. This approach has studied both individual and group fairness \cite{DBLP:conf/wsdm/RastegarpanahGC19}. In general, group fairness is easier to track and handle.

There is an abundance of machine learning algorithms used in practice for search and recommendations (and in general) dictating a clear need for a future systematic study of the relationship between dataset features, algorithms, and pre-processing performance.

%\cite{DBLP:journals/kais/KamiranC11} focus on one binary sensitive attribute and a two-class classification problem.
%\cite{DBLP:journals/corr/LumJ16}. This framework allows for an arbitrary number of variables to be adjusted and for each of these variables and the protected variables to be continuous or discrete.

 \begin{table}
\caption{Pre-processing Methods.}  \label{tab:Preprocessing}
\begin{scriptsize}
\begin{tabular}{p{2.2cm}p{0.9cm}p{0.9cm}p{1cm}p{1.5cm}}
 \hline\noalign{\smallskip}
& \textbf{bias in rows} & \textbf{bias in columns} & \textbf{fairness} & \textbf{algorithm}  \\\hline
  Suppression \cite{DBLP:journals/kais/KamiranC11} &  & \checkmark & group &  any\\\hline
    \multirow{2}{2.4cm}{Class Relabeling \cite{4909197,DBLP:journals/kais/KamiranC11}}  & \checkmark &  & group & ranker \\
    &  &  &  & \\\hline
   \multirow{2}{1.9cm}{Reweighing \cite{hainmueller2012,DBLP:journals/kais/KamiranC11,Krasanakis2018}}  &  & \checkmark &  \multirow{2}{1.7cm}{group/ individual} & ranker\\
    &  &  &  & \\\hline
    
      \multirow{2}{2.4cm}{Data transformation  \cite{DBLP:conf/nips/CalmonWVRV17} }  &  & \checkmark & \multirow{2}{1.8cm}{group/ individual}  & ranker\\
  &  &  &  & \\\hline
  
  \multirow{2}{1.9cm}{Data repair \cite{FFM+15,DBLP:journals/corr/LumJ16,DBLP:conf/sigmod/SalimiRHS19} }  & \checkmark & \checkmark  & group  &  ranker\\
    &  &  &  & \\\hline

  \multirow{2}{2.4cm}{Data augmentation \cite{DBLP:conf/wsdm/RastegarpanahGC19} }      & \checkmark & \checkmark & \multirow{2}{2cm}{group/ individual} & \multirow{2}{1.7cm}{matrix factorization}\\
 &  &  &  & \\\hline
\noalign{\smallskip}\hline
 \end{tabular}
 \end{scriptsize}
 \end{table}

\section{In-processing Methods}  \label{sec:pre-proc}
In-processing methods for achieving fairness in rankings and recommendations focus on modifying existing or introducing new models or algorithms. In this section, we survey in a unified way these methods, by distinguishing between learning approaches and approaches using preference functions.

\subsection{Learning Approaches}
For both rankings and recommenders, learning approaches typically use machine learning to construct ranking models, most often using a set of labeled training data as input. In general, the ranking model ranks unseen lists in a  way similar to the  ranking of the training data. The overall goal is to learn a model that minimizes a \textit{loss function} that captures the distance between the learned and the input ranking.

Various approaches exist, varying on the form of training data and the  type of  loss function \cite{learning-book}. In the
\textit{point-wise approach} (e.g., \cite{pointlearn1}), the training data are (item, relevance-score) pairs for each query. In this case, learning can be seen as a regression problem where given an item and a query, the goal is to predict the score of the item.
In the \textit{pair-wise} approach, the training data are pairs of items where the first item is more relevant than the second item for a given query \cite{pairlearn1,pairlearn2}. In this case, learning can be seen as a binary classification problem where given two items, 
the classifier decides whether the first item is better than the second one.
Finally, in the \textit{list-wise approach} (e.g., \cite{listnet}), the input  consists of a query and a list of items ordered by their relevance to the query.
Note that the loss function takes many different forms depending on the approach. For example, in the pair-wise approach, loss may be computed as the average number of inversions in a ranked output.

In the following, we present a number of approaches towards making the ranking models  fair.
Note that the proposed approaches can be adopted to work for different types of input data and loss functions.

%\textbf{can we put refs for point-, pair- and list-wise approaches? }

\subsubsection{Adding Regularization Terms}
A general in-processing approach  to achieving fairness is by adding \textit{regularization terms} to the loss function of the learning model.
These regularization terms express measures of unfairness that the model must minimize in addition to the minimization of the original loss function.
Depending on the  form of the training data, the loss function and the measure of fairness, different instantiations of this general approach are possible.

The DELTR approach \cite{ZDC20} extends the 
ListNet \cite{listnet} learning to rank framework.
ListNet is a list-wise framework where the
training set consists of a query $q$ and a list of items ordered by their relevance to $q$.
ListNet learns a ranking function $f$ that minimizes a loss function $L_{LN}$ that measures the extent to which the ordering $\hat{r}$ of items induced by $f$ for a query differs from the ordering $r$ that the items appear in the training set for this query. 
The loss function $L_{DELTR}$ of DELTR is: 
%\textbf{ref to ListNet?} 
\begin{multline}
L_{DELTR}(r(q), \hat{r}(q)) = L_{LN}(r(q), \hat{r}(q)) + \\
\lambda \, Unfairness(\hat{r}(q))
\end{multline}

$L_{DELTR}$ extends the original loss function $L_{LN}$ of ListNet with a term  that imposes a fairness constraint. Parameter $\lambda$  controls the trade-off between ranking utility (i.e., distance from input ranking $r$ captured by the original loss function) and fairness.

Exposure is used as a measure of unfairness in the produced output. Specifically: 
\begin{multline}
Unfairness(r(q)) = max\{0, Exposure(G^+|r(q) - \\
Exposure(G^-|r(q))^2
\end{multline}

Using the squared hinge loss makes the loss function differentiable.
Also,  the model prefers rankings in which the exposure
of the protected group is not less than the exposure of the non-protected
group but not vice versa.

A regularization approach is also taken for recommender systems in
\cite{DBLP:conf/fat/KamishimaAAS18}. Let $U$ and $I$ denote random variables for
the users and items respectively and $R$ denote a random variable for the recommendation
output. 
Let also $S$ be the sensitive attribute, that is, information to be ignored in the recommendation process, like the gender of a user or the popularity of an item. 
The goal in this case, is to achieve recommendation, or statistical, independence. This means  to include no information about the sensitive feature that influences the outcome, as well as recommendations  should satisfy a recommendation independence constraint. 

The core of this regularization approach is included in Equation \ref{eq:independence} that adopts a regularizer imposing a constraint of independence, while training the recommendation model. 
\begin{equation}\label{eq:independence}
\sum_D loss(r_i, r(x_i, y_i, s_i)) - \eta \cdot ind(R, S) + \lambda \cdot reg(\theta) 
\end{equation}
where $\eta$ is the independence parameter that controls the balance between independence and accuracy and $ind$ is the independence term, i.e., the regularizer to constrain independence; the larger value indicates that recommendations and sensitive values are more independent. Loss is the empirical loss, while $\lambda$ is the regularization parameter and $\theta$ is the L2 regularizer. 
Several alternatives can be used for the independence term, like, for example, the mutual information with histogram models or normal distributions, or by exploiting distance measures as in the case of distribution matching using the  Bhattacharyya distance.

\subsubsection{Learning via Variational Autoencoders} 

Variational Autoencoders (VAE) are proposed as the state-of-the-art for the collaborative filtering task in recommenders. With a multinomial likelihood generative model and a controlled regularization parameter, it is possible to  estimate normal distribution parameters in the middle layer of an MLP, that enriches the rating data representation and outperforms previous neural network based approaches \cite{DBLP:conf/www/LiangKHJ18}. The situation requires drawing samples from the inferred distributions in order to propagate values to the decoder, but it is not a trivial task to take gradients when having a sampling step. The \emph{re-parametrization trick}  \cite{DBLP:journals/corr/KingmaW13} is to re-parameterize
  the sampled values by incorporating a normal distributed noise, so the gradient can back-propagate through the sampled variable during the training.

Instead of using only the re-parametrization trick during the training phase,  the noise variable can be incorporated in the test phase of VAE as well \cite{DBLP:conf/medes/BorgesS19}, in order to enhance fairness in the ranking order of recommendations (using as definition of fairness, Equation \ref{eq:unfair-weikum}). The motivation here is that different noise distributions directly affect the rankings, depending on how frequently the latent values vary around the mean inside the interval defined by the variance. 
Specifically, it is experimentally shown that the noisy effect of the Gaussian and uniform distributions vary the output scores when having the same data as input, while unfairness is reduced despite of a small decrease in the quality of the ranking. The higher the variance of the new component, the greater the effect in the predicted scores, and consequently in the ranking order. 

\subsubsection{Learning Fair Representations}
The main idea in this approach is to learn a fair representation of the
input data and use it for the task at hand.
Previous work in fair classification used this idea to achieve fairness by introducing  an intermediate level $Z$ between the input space $X$ that represents individuals and the output space $Y$ that represents classification outcomes
\cite{ZWS+13}. 
$Z$ should be a fair representation of $X$ that 
best encodes $X$ and 
obfuscates any information about membership in the protected group.
Specifically, $Z$ is modeled as a multinomial random variable of size $k$ where each of the $k$ values represents a prototype (cluster) in the space of $X$. 

The goal is to learn $Z$ such that to minimize a loss function $L$:
\begin{equation}
L = \lambda_x L_x + \lambda_z L_z +  \lambda_y L_y
\label{eq:fair-rep}
\end{equation}
where the first term, $L_x$, refers to the quality of the encoding, i.e., expresses the requirement that the distance from points in $X$ to their representation in $Z$ should be small,
the second term, $L_z$, refers to fairness, and the last term, $L_y$,  refers to accuracy, i.e., the prediction based on the representation should be accurate.  Parameters $\lambda_x$, $\lambda_y$ and $\lambda_z$ are hyper-parameters that control the trade-off among these three objectives.

One can enforce different forms of fairness by appropriately defining the $L_z$ objective.
Statistical parity is used in \cite{ZWS+13} captured by the following objective:
\begin{equation}
Pr(z = k|x \in G^+) = Pr(z = k|x \in G^-), \, \forall \, k
\end{equation} 
that is, the probability that a random element that belongs to the protected group of $X$ maps to a particular prototype of $Z$ is equal to the probability that a random element that belongs to the non-protected group of $X$ maps to the same prototype.

The fair representation approach has also been used  for fair ranking instead of classification.
This was achieved by modifying the last objective 
$L_y$ in Equation \ref{eq:fair-rep} to represent accuracy in the case of ranking as opposed to accuracy in classification \cite{YS17}.
The modified objective asks that the distance between the ground truth ranking and 
the estimated ranking is small.

\subsection{Linear Preference Functions}
In some applications, items are ranked based on a score that is a weighted linear combination of the values of their attributes. Specifically, let $i$ be an item with $d$ 
 scoring attributes, $i[1] \dots i[d]$.
A linear ranking function $f$ uses
a weight vector $w$ = $(w_1, \dots w_d)$, to compute a utility (goodness) score for each item $i$,
$f(i) = \sum_{j=1}^d w_j i[j]$.

In this case, fairness is formulated as the following problem. 
Given function $f$  with weight vector
$w$ = $(w_1, \dots w_d)$, find a function $f^*$ with weight vector $w^*$ = $(w^*_1, \dots w^*_d)$, such that, $f^*$ produces a fair ranking
and its weights are as close to the weights of the original $f$ as possible, i.e.,  $cosine(w, w^*)$  is minimized \cite{AJS+19}.

\subsection{Constraint Optimization for Rank Aggregation}
For fairness-preserving rank aggregation, \cite{DBLP:journals/pvldb/KuhlmanR20} presents a solution that balances aggregation accuracy with fairness, using a pairwise rank representation. In this case, a ranking is represented as a set of pairwise comparisons between data items. Then, for two rankings, the number of pairs with items that do not have the same order in the two rankings expresses their Kendall Tau distance \cite{kendal1938}. Overall, given a set of rankings, the ranking with the minimum average Kendall Tau distance to the rankings in the set is known as the Kemeny optimal rank aggregation.

Consider the case where we have data items from two different groups, $G^+$ and $G^-$. The pairs in their Cartesian product can be divided into three subsets: pairs containing items from $G^+$, pairs containing items from $G^-$, and pairs containing one item from  $G^+$ and one item from  $G^-$. Then, $Rpar_{G^+}(\rho)$ computes the probability, for a ranking $\rho$, that an item from group $G^+$ is ranked above an item from group $G^-$: 
\begin{equation}
    Rpar_{G^+}(\rho) = P(x_i <_{\rho} x_j | x_i \in {G^+}, x_j \in {G^-})
\end{equation}

Following from this, the pairwise formulation of statistical parity for two groups is defined as: 
Given a ranking $\rho$, consisting of data items that belong to mutually exclusive groups $G^+$, $G^-$, 
$\rho$ satisfies pairwise statistical parity if the following condition is met: 
\begin{equation}
Rpar_{G^+}(\rho) = Rpar_{G^-}(\rho) 
\end{equation}

A linear integer programming solution with parity constraint is offered for aggregating many rankings, while producing a fair consensus. For achieving better efficiency, a branch-and-bound fairness-aware ranking algorithm is designed that integrates a rank parity-preserving heuristic.

~

In Section \ref{sec:summary-method}, we summarize the in-processing methods, along with the post-processing ones, highlighting the advantages of each category.

\section{Post-processing Methods}  \label{sec:post-proc}
Post-processing approaches are agnostic to the ranking or the recommendation algorithm.
Typically, they take as input a ranking $r$ and a specification of the required form of fairness and produce a new ranking $\hat{r}$ that satisfies the fairness requirements and respects the initial ranking, to the extent possible.

\subsection{Fairness as a Generative Process}
Let us consider a simple case of group parity, where we ask that a specific proportion $p$, $0 \leq p \leq 1$, of the items at the top $k$ positions in the ranking belong to the protected group.

Given $p$ and a ranking $r$, the generative process introduced in \cite{YS17}  creates a ranking $\hat{r}$ by: (a) initializing $\hat{r}$ to the empty list, and (b) incrementally adding  items to $\hat{r}$.
Specifically, for each position $j$ in $r$, a Bernoulli trial is performed with $p$. If the trial succeeds, we select the best available, i.e., most highly ranked in $r$, item from the protected group. 
Otherwise, we select the best available item from the non-protected group. An example is shown in Figure \ref{fig:ranking}. The ranking $r_l$ in the left is the original ranking based on the utility score of the items. The middle ranking $r_m$ corresponds to the extreme case of $p = 1$ where all members of the protected group are placed in the top positions, while the ranking $t_r$ on the right corresponds to $p = 0.5$. 

The produced ranking $\hat{r}$ satisfies the in-group monotonicity constraints. This means that, within each group, the items are ordered with decreasing qualifications. It is also shown that under some assumptions the ranking also maximizes the utility expressed as the average score of the items in the top-$k$ positions \cite{ZBC+17}.
 
A statistical test for this generative model is proposed in \cite{ZBC+17}. 
Specifically, given that at a specific position we have seen a specific number of items from each group, an one-tailed Binomial test is used to compare the null hypotheses that the ranking was generated using the model with parameter $p*$ = $p$, or with $p*$ $<$ $p$, which would mean that the protected group is represented less than desired.

\subsection{Fair Ranking as a Constraint Optimization Problem}
Another post-processing approach 
formulates the problem of producing a fair ranking $\hat{r}$ as an optimization problem.
Let $F$ be a fairness measure for rankings and let $U$ be a measure of the utility of a ranking for a particular task, for example, let $U(r|q)$ be
the relevance of ranking $r$ for a given query $q$.
There are two general ways of formulating an optimization problem involving fairness $F(r)$ and utility $U(r|q)$, namely: 
\begin{itemize}
\item ({\sc maxFcoU}) maximizing fairness subject to a constraint in utility,
\item ({\sc maxUcoF}) maximizing utility subject to a constraint in fairness.
%and, \item maximizing a combination of utility and fairness ({\sc hybrid}).
\end{itemize}

In the {\sc maxFcoU} formulation, the underlying idea  is to produce a ranking   that is as fair as possible while remaining   
 relevant to $q$ (e.g., \cite{BGW18}). 
For example, we ask for the most fair ranking $\hat{r}$ among all rankings such that $\hat{r}$ also satisfies a utility constraint, e.g., the  loss in utility with regards to the original ranking $r$ remains below a given threshold $\theta$:

$\hat{r} = argmax_{\hat{r}}F(\hat{r})$ \\
\hspace*{0.12in} s. t., $  distance(U(r|q), U(\hat{r}|q)) \leq \theta$.

Alternatively, in the {\sc maxUcoF} formulation, we  look for the ranking  that has the
the maximum possible utility among all rankings whose fairness is sufficient
(e.g., \cite{BGW18}, \cite{SJ18}). For example, the approach  proposed in \cite{SJ18} produces a ranking $r$ such that:

$r = argmax_r U(r|Q)$  
\hspace*{0.12in} s.t. $r$ is fair.

We characterize such {\sc maxUcoF} approaches as post-processing, since they assume that the utility of each item is known, or can be estimated. Thus, implicitly there is an original, non-fair ranking in which the items are  ordered solely by their utility. Then, given these individual utilities, a new ranking is produced that also satisfies a fairness constraint.
%% --

In general, the complexity of both the {\sc maxFcoU} and {\sc maxUcoF} optimization problems depends on the type of the utility and fairness functions, and the form of the constraints. In some cases, the optimization problems can be solved using linear integer programming (ILP), e.g., \cite{BGW18}, or, in some special cases using a dynamic programming algorithm  e.g., \cite{CSV18}.

Constraints can be also used for  producing fair recommendation packages of items for groups of users \cite{DBLP:conf/www/SerbosQMPT17}. The intuition of the method is to greedily construct  a package $P$, by adding in rounds to $P$ the item that satisfies the largest number of non-satisfied users. Specifically, given that $sat_G(P)$ denotes the users in $G$ that are satisfied by $P$, at each round, the goal is to maximize: 
\begin{equation}
f_G(P,i) = |sat_G(P \cup {i}) \backslash sat_G(P)|. 
\end{equation}

When considering fairness proportionality (Equation \ref{eq:fenvy}), $sat_G(i)$ contains the users for which item $i$ belongs in their top-$\delta$\% most preferable items. For envy-freeness fairness (Equation \ref{eq:fprop}), $sat_G(i)$ contains the users that are envy-free for the item $i$. That is, $P$ is fair for a user $u$ (m-proportional, or m-envy-free), if there are at least $m$ items $i$ in $P$, such that, $u \in sat_G(i)$.

This method is generalized to include constraints that restrict the set of candidate packages that can be recommended to a group of users. Two types of constraints are discussed, namely \textit{category} and \textit{distance} constraints. In simple words, with category constraints, when selecting an item from a specific category, we remove the items of this category from the candidate set. With distance constraints, we consider as candidate items only the items that when added to the existing solution satisfy the specific input distance constraints.

\subsection{Fairness via Calibration Methods}
Calibration methods suggest to re-rank a list of  items as a post-processing step. 
Equation \ref{eq:kl} (in Section \ref{sec:fair-recs}) quantifies the degree of calibration for recommender's outputs
 based on specific metrics    \cite{DBLP:conf/recsys/Steck18}.
 
To determine the optimal set $I^*$ of $N$ recommended items -- movies in the suggested scenario -- the maximum marginal relevance function \cite{DBLP:journals/sigir/CarbinellG17} is used: 
\begin{equation}\label{eq:klmax}
I^* = argmax_{I, |I|=N} (1- \lambda) \cdot s(I) - \lambda \cdot C_{KL}(p,q(I))
\end{equation} 
where $\lambda \in [0,1]$ determines the trade-off between the prediction scores $s(i)$ of the movies $i \in I$, with $s(I) = \sum_{i \in I} s(i)$, and the calibration metric (Equation \ref{eq:kl}). 
Namely, the trade-off between accuracy and calibration is controlled by $\lambda$. Greedily, the method starts with an empty set, and iteratively adds one movie at a time, namely the movie that maximizes Equation \ref{eq:klmax}. 

% targets at locating the best suggestions $I$ for a group $G$ that maximize the social welfare $SW(G,I)$ (Equation \ref{eq:sw}) and fairness $F(G,I)$ (Equation \ref{eq:fsw1} and \ref{eq:fsw2}), using the scheme: 
In a similar way, for group recommendations, the best suggestions $I$ for a group $G$ should maximize the social welfare $SW(G,I)$ (Equation \ref{eq:sw}) and fairness $F(G,I)$ (Equation \ref{eq:fsw1} and \ref{eq:fsw2}), using the scheme \cite{DBLP:conf/recsys/LinZZGLM17}: 
\begin{equation}
\lambda \cdot SW(G,I) + (1-\lambda) \cdot F(G,I). 
\end{equation}

A greedy solution is to select an item that, when added to the current recommendation list, achieves the highest fairness $F(G,I)$. More time-efficient alternatives are offered via integer programming techniques.

When considering the notion of Pareto optimality for group recommendations, a simple heuristic can be used to compile and approximately identify the list of the top-$N$ recommendations for a group \cite{DBLP:conf/sac/Sacharidis19}:
 Specifically, given $N'$, where $N'>N$ is the largest number of items the system can recommend for an individual user, the method proceeds as follows: (a) it requests the top-$N'$ recommendations for each user in the group, (b) it takes their union, and (c) it identifies the $N$-level Pareto optimal items among the items in the union.

\subsection{Fairness in Multiple Outputs} 
\label{sec:seq-recs-fairness}

When providing rankings or recommendations, users typically pay more attention to the first positions, and attention wears off for items in lower positions in the ranking. In a situation where the greatest estimated probabilities are quite close or equal to each other, the algorithm needs to arrange them in a proper order and necessarily present high scores in high positions. This promotes an unfair result that can be mitigated in the long term, by changing the position of items in sequential rounds of rankings or recommendations. 

In the case of rankings, one approach  \cite{BGW18} is to require that ranked items receive attention that is proportional to their utility in a sequence of rankings (Equation \ref{eq:equity}).
This way, the unfair position one item appears in a single ranking can be compensated in the next rankings when it changes position, and the whole session contemplates long-term fairness. However, the  act  of  reducing  unfairness  implies reduction in the ranking quality, due to the perturbation of the utility-based rankings. 
The trade-off between ranking quality and fairness is formulated as a constrained optimization problem \cite{BGW18}
targeting at minimizing unfairness subject to constraints on quality (i.e., lower-bound the minimum acceptable quality). 

%($R$) distributions with constraints on the NDCG-quality loss in each ranking. Formally: 
Specifically, for a sequence $\rho^1 \dots \rho^m$ of rankings where the items are ordered by the utility score, inducing zero quality loss, the aim is at reordering them into $\rho^{1*} \dots \rho^{m*}$, to minimize the distance between the attention ($A$) and utility ($U$) distributions with constraints on the NDCG-quality loss in each ranking. Formally: 
\begin{equation}
    min \sum_i |A_i - U_i|, 
\end{equation}
subject to $\text{NDCG-quality@k}(\rho^j,\rho^{j*}) \geq \theta, j=1,\ldots, m$,\\
where $A_i$ and $U_i$ denote the cumulative attention and utility scores that the item $i$ gained across all rankings.

Consider  a different scenario, where a group of users, interacts with a recommender multiple times \cite{DBLP:conf/sac/StratigiNPS20}.
When following traditional methods for group recommendations, like the average aggregation method and the least misery one, the degree of \textit{satisfaction} for each user in the group (e.g., Equation \ref{eq:sequentialsatisfaction}), can not be good enough for all users in the group, leading to cases in which almost none of the reported items are of interest to some users in the group. That is, the recommender system is unfair to these users, and unfairness continues throughout a number of recommendations rounds. 
To overcome the drawbacks of the \textit{average} and the \textit{least misery} aggregation methods, and capitalize on their advantages,
an aggregation method, called \textit{sequential hybrid aggregation method},  offers a weighted combination of them \cite{DBLP:conf/sac/StratigiNPS20}. Specifically: 
\begin{multline}
score(G,d_z,j) = (1-\alpha_j)*avgScore(G,d_z,j) + \\  \alpha_j*leastScore(G,d_z,j).    
\end{multline}

For a group $G$, $avgScore(G,d_z,j)$ returns the score of the item $d_z$ as it is computed by the average aggregation method during round $j$, and $leastScore(G,d_z,j)$ returns the least satisfied user's score of $d_z$ at round $j$. 
To self-regulate the value of $\alpha$ between 0 and 1, so as to more effectively describe the consensus of the group, $\alpha$ is set dynamically in each iteration by subtracting the minimum satisfaction score of the group members in the previous iteration, from the maximum score:  
\begin{multline}
\alpha_j = max_{u \in G}sat(u,Gr_{j-1}) - \\ min_{u \in G}sat(u, Gr_{j-1}), 
\end{multline}
where $sat(u,Gr_{j-1})$ defines the satisfaction of user $u$ for the group recommendations $Gr_{j-1}$ at round $j-1$. 

The dynamic calculations of $\alpha$ counteracts the individual drawbacks of the average and least misery method. Intuitively, if the group members are equally satisfied at the last round, then $\alpha$ takes low values, and the aggregation will closely follow that of average, where everyone is treated as an equal. On the other hand, if one group member is extremely unsatisfied in a specific round, then $\alpha$ takes a high value and promotes that member's preferences on the next round. 

Next, we summarize both the in-processing and post-processing methods, and provide their advantages.

\section{Summary of In- and Post-processing Methods}  \label{sec:summary-method}

In this section, we summarize the in-processing and post-processing approaches for achieving fairness. Overall, there exist methods that have been proposed in the context of rankings, recommender systems, as well as for the rank aggreagation problem. Table \ref{tab:InPostprocessing} organizes the methods based on (a) the level of fairness, namely, individual or group, (b) the side of fairness, namely, consumer or producer, and (c) the output multiplicity, namely, if a method focuses on a single or multiple outputs. 

In general, all existing \textit{learning} and \textit{linear preference functions} in-processing approaches target group and producer fairness. Most approaches consider a single output with the recent exception of \cite{DBLP:conf/medes/BorgesS19} using VAEs that considers multiple outputs. 
Typically, the approaches in this category extend the objective function that they use by including a fairness, or unfairness, term. The target is to find the best balance between the accuracy objective and fairness objective of the optimization problem that a particular approach applies.

Regarding the post-processing approaches, we observe that there exist works focusing on all different options of fairness definitions. 
As in the pre-processing case, post-processing methods treat the algorithms for producing rankings and recommendations as black boxes, without changing their inner workings. 
This means that any post-processing approach receives as input a ranked output, and re-ranks the data items in this output to improve fairness, respecting the initial ranked output, to the extent possible. 

Typically, in-processing approaches manage to offer better trade-offs between fairness and accuracy, compared to the post-processing methods, since they naturally identify this balance via the objective function they use. However, at the same time, they cannot offer any guarantees about fairness in the output rankings and recommendations, since fairness is considered only during the training phase. 
When considering post-processing approaches, it is important to note that they can lead to unpredictable losses in accuracy, since they treat the algorithms for producing rankings and recommendations as black boxes. This is especially true for pre-processing methods as well. 
On the positive side, post-processing methods offer outputs that are easy to understand, when comparing their outputs with the outputs before any application of a post-processing fairness method, and realize that they offer a fairer output.

Recently, \cite{ZWS+13} combines both pre-processing and in-processing strategies by jointly learning a
fair representation of the data and the classifier parameters. This approach has two main limitations: (a) it leads to a non-convex optimization problem and does
not guarantee optimality, and (b) the accuracy of the classifier depends on the dimension of the fair representation, which needs to be chosen rather arbitrarily.

%Many of the prior studies are restricted to a narrow range of classifiers, and they only accommodate a single, binary sensitive attribute \cite{DBLP:journals/jmlr/ZafarVGG19}.
%, and (iii) they cannot eliminate disparate treatment and disparate impact simultaneously \cite{DBLP:journals/jmlr/ZafarVGG19}.

\begin{table*}[t]
\caption{In- and Post-processing Methods.} 
\label{tab:InPostprocessing}
\begin{tabular}{|l||l|l|l|l|l|l|}
\hline 
\multirow{2}{*}{ } & \multicolumn{2}{|l|}{\textbf{Level of fairness}} & \multicolumn{2}{|l|}{\textbf{Side of fairness}} & \multicolumn{2}{|l|}{\textbf{Output multiplicity}} \\ \hline
 & Individual & Group & Consumer & Producer & Single & Multiple \\ \hline \hline
\multicolumn{7}{|l|}{\textbf{In-processing methods}}                                \\ \hline \hline
Adding regularization terms & & \cite{ZDC20,DBLP:conf/fat/KamishimaAAS18} & & \cite{ZDC20,DBLP:conf/fat/KamishimaAAS18} & \cite{ZDC20,DBLP:conf/fat/KamishimaAAS18} &                                                           \\ \hline
Learning fair representations & & \cite{ZWS+13} & & \cite{ZWS+13} & \cite{ZWS+13} &                                                           \\ \hline
Learning with VAEs & & \cite{DBLP:conf/medes/BorgesS19} & & \cite{DBLP:conf/medes/BorgesS19} & & \cite{DBLP:conf/medes/BorgesS19} \\ \hline
Linear preference functions & & \cite{AJS+19} & & \cite{AJS+19} & \cite{AJS+19} & \\ \hline 
Constraint optimization for rank aggregation & & \cite{DBLP:journals/pvldb/KuhlmanR20} & \cite{DBLP:journals/pvldb/KuhlmanR20} & & \cite{DBLP:journals/pvldb/KuhlmanR20} & \\ \hline \hline
\multicolumn{7}{|l|}{\textbf{Post-processing methods}}                                \\ \hline \hline
Fairness as a generative process              &                                                                                & \cite{YS17,ZBC+17}                                                 &                   &  \cite{YS17,ZBC+17}                 & \cite{YS17,ZBC+17}                                                             &                                                           \\ \hline
Fairness as a constraint optimization problem & \cite{DBLP:conf/www/SerbosQMPT17,BGW18}                       & \cite{SJ18,CSV18}               & \cite{DBLP:conf/www/SerbosQMPT17}                             & \cite{SJ18,BGW18,CSV18}         & \cite{DBLP:conf/www/SerbosQMPT17,SJ18,CSV18}                                           & \cite{BGW18}                             \\ \hline
Fairness with calibration methods             & \cite{DBLP:conf/sac/Sacharidis19,DBLP:conf/recsys/LinZZGLM17} & \cite{DBLP:conf/recsys/Steck18} & \cite{DBLP:conf/sac/Sacharidis19,DBLP:conf/recsys/LinZZGLM17} & \cite{DBLP:conf/recsys/Steck18} & \cite{DBLP:conf/sac/Sacharidis19,DBLP:conf/recsys/Steck18,DBLP:conf/recsys/LinZZGLM17} &                                                           \\ \hline
Fairness in multiple rounds                   & \cite{DBLP:conf/sac/StratigiNPS20,BGW18}                      &                                                  & \cite{DBLP:conf/sac/StratigiNPS20}                            & \cite{BGW18}                    &                                                                                                         & \cite{DBLP:conf/sac/StratigiNPS20,BGW18} \\ \hline
\end{tabular} 
\label{tab:fairmethodstaxonomy2}
\end{table*}

\section{Verifying Fairness}  \label{sec:fair-progr}
In the previous sections,  we studied methods for achieving fairness. Both pre-processing and post-processing methods treat the recommendation or ranking algorithm as a black box and try to address fairness in the input or the output of the algorithm, respectively. In particular, post-processing methods assume that we already know that we have an algorithm that creates discrimination in its output and try to mitigate that. The question that naturally arises is how  we can verify whether a program is fair in the first place. 

\emph{Program fairness verification} aims at analyzing a given decision-making program and constructing a proof of its fairness or unfairness — just as a traditional static program verifier would prove correctness of a program with respect to, for example, lack of divisions by zero. However, there are several challenges:
(a) what class of decision-making programs the program model will capture,  
(b) what  the input to the program is,
(c) how to describe what it means for the program to be fair, and
(d) how to fully automate the verification process.

One simple approach is to take a decision-making program $P$ and a dataset as input \cite{DBLP:conf/sigsoft/GalhotraBM17}. Using a concrete dataset simplifies the verification problem but it also raises questions of whether the dataset is representative for the population for which we are trying to prove fairness. An alternative approach would be to use a population model $M$ as input, which can be a probabilistic model that defines a joint probability distribution on the inputs of $P$ \cite{DBLP:journals/corr/AlbarghouthiDDN16}. 
Then, we need to define when and why a program is fair or unfair. If we want to prove group fairness — for example, that the algorithm is just as likely to hire a minority applicant ($m$) as it is for other, non-minority applicants we could define a post-condition like the following \cite{DBLP:journals/corr/AlbarghouthiDDN16}:

$\frac{Pr[P(v) = true | v_s =m]}{Pr[P(v) = true | v_s \neq m]}>1 - \epsilon$ \\

The verifier then will prove or disprove that $P$ is fair for the given population.

%In this section, we examine a more fundamental question: fairness as a program property. There are two approaches: (a) verifying if a program is fair, and (b) having fairness as a first-class citizen in programming.

%A \emph{fairness verifier} takes a decision-making program $P$ and a dataset or a population model $M$ as input. While using a concrete dataset simplifies the verification problem \cite{DBLP:conf/sigsoft/GalhotraBM17}, it also raises questions of whether the dataset is representative for the population for which we are trying to prove fairness. A population model $M$ can be a probabilistic model that defines a joint probability distribution on the inputs of $P$ \cite{DBLP:journals/corr/AlbarghouthiDDN16}.

%\begin{figure}
%  \includegraphics[width=\linewidth]{figures/verifier.jpg}
%\caption{A program verifier}
%\label{fig:verifier}       % Give a unique label
%\end{figure}

In general, proving group fairness is easier: the verification process reduces to computing the probability of a number of events with respect to the program and the population model. However, proving individual fairness requires more complex reasoning involving multiple runs of the program, a notoriously hard problem. Moreover, in the case of a negative result, the verifier should provide the users with a proof of unfairness. Depending on the fairness definition, producing a human readable proof might be challenging as the argument might involve multiple and potentially infinite inputs. 
For example, for group fairness, it might be challenging to explain why the program outputs true on $40\%$ of the minority inputs and on $70\%$ of the majority inputs. Overall, program fairness verification is a difficult, and less investigated, topic. 

A different approach is to make fairness a first-class citizen in programming. 
In \emph{fairness-aware programming} \cite{DBLP:conf/fat/AlbarghouthiV19}, developers can state fairness expectations natively in their code, and have a run-time system monitor decision-making and report violations of fairness.
This approach is analogous to the notion of assertions in  modern programming languages. For instance, the developer might assert that $ x > 0$, indicating that they expect the value of $x$ to be positive at a certain point in the code. 
The difficulty, however, is that fairness definitions are typically probabilistic, and therefore detecting their violation cannot be done through a single execution as in traditional assertions. Instead, we have to monitor the decisions made by the procedure, and then, using statistical tools, infer that a fairness property does not hold with reasonably high confidence.

For example, consider a movie recommendation system, where user data has been used to train a recommender that, given a user profile, recommends a single movie \cite{DBLP:conf/fat/AlbarghouthiV19}.
Suppose that the recommender was constructed with the goal of ensuring that male users are not isolated from movies with a strong female lead. 
Then, the developer may add the following specification to their recommender code:

$@spec(pr(femaleLead(r)|s = male) > 0.2)$

The above specification ensures that for male users, the procedure recommends a movie with a female lead at least $20\%$ of the time.

To determine that a procedure $f$ satisfies a fairness specification $\phi$, we need to maintain statistics over the inputs and outputs of the procedure $f$ as it is being applied. Specifically, we compile the specification $\phi$ into run-time monitoring code that executes every time $f$ is applied, storing aggregate results of every probability event appearing in $\phi$. 
In the earlier example with movie recommendation, the monitoring code would maintain the number of times the procedure returned true for a movie with a female lead. 
Again, a big challenge is checking individual fairness. In this case, the run-time system has to remember all decisions made explicitly, so as to compare new decisions with past ones. 

%\cite{DBLP:conf/aies/CortesG20}, \cite{DBLP:conf/fat/KallusMZ20}

%Clearly, both program fairness verification and fair-aware programming are novel approaches that need further development.

\section{Open Challenges} \label{sec:challenges} 
In this paper, we have just begun to realize the need for fairness in particular fields, namely in rankings and recommender systems. 
Next, we highlight a few critical open issues and challenges for future work, which aim to support advanced services for making accountable complex rankings and recommender systems. 

\paragraph{A codification of definitions.}
As described in Sections \ref{sec:fair-problem} and \ref{sec:fair-rank-recs}, there is not a universal definition for expressing fairness in rankings and recommenders. Instead, the list of potential definitions is very long. 

Specifically, just for fairness alone, there exist a plethora of different definitions. Some of them are not even compatible, in the sense that there is no method that can satisfy all of them simultaneously, except in highly constrained special cases \cite{KMR17}. Furthermore, there is also a need to make explicit the correspondence between these definitions and the interpretation of bias or diversity, that each of them materializes. The limitations of each definition, the compatibility among them, the incurred trade-offs, the domain of applicability, assumptions and parameters are not well understood yet. 

Moreover, it is interesting to see how the general public views fairness in decision making. By testing people’s perception of different fairness definitions, we can understand definitions of fairness that are appropriate for particular contexts \cite{DBLP:conf/www/Grgic-HlacaRGW18},\cite{DBLP:conf/uss/PlaneRMT17}. One such attempt investigates which definitions people perceive to be the fairest in the context of loan decisions \cite{DBLP:conf/aies/SaxenaHDRPL19}, and whether fairness perceptions change with the addition of sensitive information (i.e., race of the loan applicants). 
%Overall, one definition (calibrated fairness) tended to be more preferred than the others. 

\paragraph{Lack of data.} A major challenge is that the available data is often limited \cite{DBLP:conf/icde/AsudehJJ19}. This way, any analysis is done with data that has been acquired independently, through a process on which the data scientist has limited control. Collecting more data for analysis is challenging, and will help to discover more types of biases on it. 
In the long run, one could envision benchmarks for measuring the societal impact of an algorithm along the lines of the TPC benchmark\footnote{http://www.tpc.org/information/benchmarks.asp} for database performance. 

\paragraph{A unified approach for the data pipeline.}
One limitation of the current work is that fairness has been studied for specific tasks in isolation, with most current work in fairness focusing on the classification task with the goal of non-discrimination. However, there is a need to consider fairness along the whole data pipeline \cite{DBLP:conf/ssdbm/StoyanovichHAMS17}. This includes pre-processing steps, such as data selection, acquisition, cleaning, filtering and integration. Pre-pro- cessing for removing bias, can be viewed as the action of repairing, e.g., by replacing, modifying, or deleting data that cause bias. 
This pipeline also includes post-processing steps, such as data representation, data visualization and user interfaces. 
For example, how results are presented can introduce bias, and this is why we need to understand the implications for fairness. 

\paragraph{Lack of evaluation tools.}
Besides coming up with the correct way of defining fairness in rankings and recommenders, there is also a need for tools for investigating bias, and evaluating the quality of a dataset, an algorithm, or a system. There are some first attempts, such as IBM’s AI Fairness 360\footnote{ https://aif360.mybluemix.net/} and Tensor Flow’s Fairness Indicators\footnote{https://www.tensorflow.org/tfx/guide/fairness\_indicators}. However, both of them focus mainly on statistical group measures of fairness in classification. Concepts, like context and provenance, are important and can be directly considered in designing such tools. 

In this direction, we also need efficient ways for measuring fairness and monitoring its evolution over time. Previous research in stream processing and incrementally maintaining statistics may be relevant here.

Perhaps the most pending question is how to quantify the long-term impact of enforcing methods that target at ensuring fairness. Would they work in favor of the social good, or would they backfire in ways that we cannot predict?

\paragraph{Lack of real applications of fairness.} 
While a lot of work is done in a research setting, we still do not see many actual applications and their results. There are many challenges in making algorithms and systems fairer in the real-world. A company, for example, needs to consider its business metrics (e.g., click-through rate, purchases) and make sure that these are not affected. For example, how to design algorithms that take into account all these different objectives is challenging. 

An example of a real application of fairness is found in LinkedIn \cite{DBLP:conf/kdd/GeyikAK19}, where they developed a fair framework and they applied it to LinkedIn Talent Search. 
Online A/B tests showed considerable improvement in the fairness metrics  without a significant impact on the business metrics, which paved the way for deployment to LinkedIn users worldwide.

\paragraph{A multi-level architecture of value systems and algorithms.} 
A problem intrinsic to the definition of all fairness definitions in rankings and recommenders is the fact that they attempt to quantify philosophical, legal, often elusive, and even controversial notions of justice and social good. Complexity is aggregated when notions that reflect value systems and beliefs interact with the mechanisms for implementing them. In the very least, there should be a clear distinction between what constitutes a belief and what is the mechanism, or measure, for codifying this belief. From a technical point of view, we should then be able to focus on assessing whether a proposed measure is an appropriate codification of a given belief as opposed to assessing the belief itself.

This calls for developing different levels of abstractions and mappings between them. This is somehow reminiscent of how data independence is supported in database management systems by the three-level architecture with the physical, conceptual, and external level and the mappings between these levels \cite{DBLP:books/daglib/0011128}. At the lower level, we could have beliefs and value systems and at the higher level fairness definitions. Intermediate levels could be used to support transformations for getting from the lower to the higher level.

\paragraph{Relating algorithmic fairness with other notions of fairness in systems.} 
The focus of this survey is on fairness in decision-making processes, and in particular on rankings and  recommendation systems used in this context.
However, there are several other cases where a system needs to make decisions, and where fairness is also important. In particular, there are several problems more familiar to the data management community, such as resource allocation and scheduling, where fairness has been studied. We present here representative examples.

For instance, an approach is presented 
in \cite{DBLP:conf/sigmod/KunjirFMB17} for cache allocation where fairness is based on Pareto efficiency and sharing incentives.
A multiple resource allocation approach is introduced in \cite{DBLP:conf/nsdi/GhodsiZHKSS10} that generalizes max-min fairness to multiple resource types, where max-min fairness in this context refers to maximizing the minimum allocation received by a user in the system. 
Another notion of fairness, termed proportionate progress, is proposed in \cite{DBLP:journals/algorithmica/BaruahCPV96} for the periodic scheduling problem where weighted resources are allocated to tasks for specific time units in each interval. In proportionate progress fairness, each task is scheduled resources according to its importance (i.e., its weight). 
Finally, for scheduling,  in \cite{DBLP:conf/edbt/GuptaMWD09}, fairness means that the workload scheduler performs in a way that no query starves for resources.

Another notion of fairness, for a different problem, namely the chairman selection problem,  is proposed in \cite{DBLP:journals/dm/Tijdeman80}.
In the chairman selection problem, a set of states want to form a union and select a chairman for each year. Fairness in this context refers to guaranteeing a small discrepancy for the number of chairmen representing each state, so that all states are satisfied. 

Clearly, fairness concerns have emerged in various contexts through the years. Providing a unified view of all these different notions of fairness and the computational methods used to enforce them is an open problem and an opportunity. How to embed models and algorithms for fairness into any system that involves any type of decision making is also open. Of course, new models and algorithms may be needed in different contexts, where the principles of fairness may be different. What works well for a recommendation problem may not work well for a query optimizer or resource sharing in cloud computing. However, 
this unified view will potentially provide new insights and opportunities for cross-fertilization.

\paragraph{Fairness in other domains.}  
In this paper, we have focused on ranking and recommendation algorithms.
Many more algorithms are being revisited under the lens of fairness. Two such examples are clustering and ranking in networks.
For instance, fair clustering adopts a parity definition of fairness and asks that
each group must have approximately equal representation in every cluster \cite{fairlets}. 
For the link analysis problem in networks, fair algorithms  are introduced in \cite{DBLP:conf/www/Tsioutsiouliklis21} that use a parity-based definition of fairness and apply constraints on the proportion of Pagerank allocated to the members of each group of nodes in the network. 

Algorithmic fairness is a fast changing field. It is an open challenge to provide fairness measures and definitions for all kinds of algorithms and use them in evaluating such algorithms, in analogy of, say response time, for measuring  performance \cite{pit20}.

Recently, the concept of fairness has also been studied in different domains and for covering different needs. We discuss two such domains, labor market and social matching.
For instance, previous research examines  labor market cases under fairness concerns and shows the relevance of such concerns on economic outcomes especially when considering employment contracts over time \cite{AnnualReviewEconomics09}. 
Recently, in the same domain, the work in \cite{HuC18}  pinpoints the persistence of racial inequalities and designs solutions with respect to a dynamic reputational model of the labor market, highlighting the results from groups divergent accesses to resources.
Other recent work examines the fairness of online job marketplaces  in terms of ranking job applicants \cite{sihem}. Instead of partitioning the
individuals in predefined groups, the authors seek to
find a partitioning of the individuals based on their protected
attributes that exhibits the highest unfairness.

For the problem of professional social matching,  conventional mechanisms, such as optimizing for similarity and triadic closure, are studied in \cite{DBLP:journals/cacm/0002HK20} and shown to  involve risks of strengthening the homophily bias and echo chambering. For the problem of team assembly, the work in \cite{DBLP:conf/webi/MachadoS19} designs fairness-aware solutions when multidisciplinary teams need to be formed and allocated to work on different projects with requirements on members' skills.

The application of  models and mechanisms   of algorithmic fairness in a variety of domains that involve social and economic activities, such as in labor market, social matching and team formation, opens a lot of opportunities for fruitful interdisciplinary work.

\section{Conclusions}  \label{sec:conclusions}
Ranking and recommender systems have several applications, such as hiring, lending, and college admissions, where the notion of fairness is very important since decision making is involved. 
We have only begun to understand the nature, representation and variety of the several definitions of fairness, and the appropriate methods for ensuring it.

In this article, we follow a systematic and structured approach to explain the various sides of and approaches to fairness. First, we lay the ground by presenting general fairness definitions. Then, we zoom in on models and definitions for rankings, recommendations and the problem of rank aggregation. We organize them in a taxonomy and highlight their differences and commonalities. %, as well as the opportunities for cross-domain transfer. 
This analysis naturally leads to a number of open questions, like: (a) How do fairness definitions fare? (b) Which definition is suitable for which context? (c) How do people perceive fairness in different contexts? (d) What does it mean to be fair after all? Is there a unified way to be able to judge whether an outcome or an algorithm is fair?

We move on to describing solutions for fair rankings and recommendations. We organize the approaches to tackle unfairness or ensure a fairer outcome into pre-, in- and post-processing approaches. Within each category, we further classify them along several dimensions. 
%We have seen that approaches to tackle unfairness or ensure a fairer outcome fall into different categories, and within each category, different approaches exist. 
It is still very early to say which one works best for which context. There is no evaluation that puts them all under the same lens and there are generally not conclusive results as to which fare better. It may be the case that a combination of methods should be applied, e.g., combining pre-processing and in-processing steps. This is also an open question as different efforts have adapted a single angle in the problem.

While the focus of this survey is on fairness in rankings and recommendation systems, we discuss several other cases where a system needs to make decisions and fairness is also important, and how we can verify whether a program is fair. 
Finally, we discuss open research challenges pertaining to fairness in the broader context of data management and on designing, building, managing, and evaluating fair data systems and applications.

%In this context, it is interesting to see how the general public views fairness criteria in algorithmic decision making.

%\begin{figure}
%  \includegraphics{example.eps}
%\end{figure}

% \begin{figure*}
%   \includegraphics[width=0.75\textwidth]{example.eps}
% \caption{Please write your figure caption here}
% \label{fig:2}       % Give a unique label
% \end{figure*}
 
% \begin{table}
% \caption{Please write your table caption here}
% \label{tab:1}       % Give a unique label

% \begin{tabular}{lll}
% \hline\noalign{\smallskip}
% first & second & third  \\
% \noalign{\smallskip}\hline\noalign{\smallskip}
% number & number & number \\
% number & number & number \\
% \noalign{\smallskip}\hline
% \end{tabular}
% \end{table}

%\begin{acknowledgements}
%\end{acknowledgements}


\bibliographystyle{spmpsci}
\bibliography{fairRR}

\end{document}